\pdfoutput=1

\documentclass[10pt, conference, compsocconf]{IEEEtran}

\let\labelindent\relax

\usepackage[normalem]{ulem}
\usepackage[usenames,dvipsnames,svgnames,table]{xcolor}
\usepackage{graphicx}
\usepackage{listings}
\usepackage{lstlinebgrd}
\usepackage{epstopdf}
\usepackage{subcaption}
\usepackage{xspace}
\usepackage{latexsym}
\makeatletter
\newif\if@restonecol
\makeatother

\usepackage[ruled,vlined,linesnumbered]{algorithm2e}
\usepackage{algorithmic}
\usepackage{enumerate}

\usepackage{xcolor-solarized}
\usepackage{textcomp}
\usepackage{wrapfig}
\usepackage{layouts}
\usepackage{setspace}
\usepackage{blindtext}
\usepackage{amsfonts}
\usepackage{multirow}
\usepackage{pifont}
\usepackage{mathrsfs}
\usepackage{mathtools}
\usepackage{relsize}
\usepackage{ulem}
\usepackage{comment}
\usepackage{footnote}
\usepackage{enumitem}
\usepackage{xspace}
\makesavenoteenv{tabular}
\makesavenoteenv{table}
\usepackage{color}

\usepackage{caption}
\DeclareCaptionType{copyrightbox}

\newif{\ifSubmit}
\newif{\ifFinal}
\newif{\ifDraft}
\Drafttrue

\newlength{\gapspace}

\newcommand{\proj}{\textsc{Wukong}\xspace}

\usepackage{booktabs}
\usepackage{siunitx}

\usepackage{cleveref}
\crefformat{section}{\S#2#1#3}
\crefname{figure}{figure}{figures}
\crefname{table}{table}{tables}

\usepackage{url}
\usepackage{balance}

\usepackage[square,numbers]{natbib}
\usepackage{xifthen}
\newcommand{\myparagraph}[2][]{\ifthenelse{\isempty{#1}}{\vspace{0.02in}\noindent\textbf{#2}}{\vspace{-0.02in}\noindent\textbf{#2}}}

\colorlet{punct}{red!60!black}
\definecolor{background}{HTML}{EEEEEE}
\definecolor{delim}{RGB}{20,105,176}
\colorlet{numb}{magenta!60!black}

\ExplSyntaxOn
\NewDocumentCommand \lstcolorlines { O{green} m }
{
 \clist_if_in:nVT { #2 } { \the\value{lstnumber} }{ \color{#1} }
}
\ExplSyntaxOff

\lstdefinelanguage{json}{
    basicstyle=\normalfont\ttfamily,
    numbers=left,
    numberstyle=\scriptsize,
    stepnumber=1,
    numbersep=8pt,
    showstringspaces=false,
    breaklines=true,
    frame=lines,
    backgroundcolor=\color{background},
    literate=
     *{0}{{{\color{numb}0}}}{1}
      {1}{{{\color{numb}1}}}{1}
      {2}{{{\color{numb}2}}}{1}
      {3}{{{\color{numb}3}}}{1}
      {4}{{{\color{numb}4}}}{1}
      {5}{{{\color{numb}5}}}{1}
      {6}{{{\color{numb}6}}}{1}
      {7}{{{\color{numb}7}}}{1}
      {8}{{{\color{numb}8}}}{1}
      {9}{{{\color{numb}9}}}{1}
      {:}{{{\color{punct}{:}}}}{1}
      {,}{{{\color{punct}{,}}}}{1}
      {\{}{{{\color{delim}{\{}}}}{1}
      {\}}{{{\color{delim}{\}}}}}{1}
      {[}{{{\color{delim}{[}}}}{1}
      {]}{{{\color{delim}{]}}}}{1},
}

\begin{document}

\title{
In Search of a Fast and Efficient Serverless DAG Engine
}


\author{
	{\rm Benjamin Carver, Jingyuan Zhang, Ao Wang, Yue Cheng}\\
	{{\it George Mason University}}
}

\maketitle


\pdfoutput=1

\begin{abstract}
Python-written data analytics applications can be modeled as and compiled into a directed acyclic graph (DAG) based workflow, where the nodes are fine-grained tasks and the edges are task dependencies. 
Such analytics workflow jobs are increasingly characterized by short, fine-grained tasks with large fan-outs. These characteristics make them well-suited for a new cloud computing model called serverless computing or Function-as-a-Service (FaaS), which has become prevalent in recent years. The auto-scaling property of serverless computing platforms accommodates short tasks and bursty workloads, while the pay-per-use billing model of serverless computing providers keeps the cost of short tasks low.

In this paper, we  thoroughly investigate the  problem space of DAG scheduling in serverless computing. We identify  and  evaluate  a  set  of  techniques to make DAG schedulers serverless-aware. These techniques have been implemented in {\proj}, a serverless, DAG scheduler attuned to AWS Lambda. {\proj} provides decentralized scheduling through a combination of static and dynamic scheduling. We present the results of an empirical study in which {\proj} is applied to a range of microbenchmark and real-world DAG applications. Results demonstrate the efficacy of {\proj} in minimizing the performance overhead introduced by AWS Lambda --- {\proj} achieves competitive performance compared to a serverful DAG scheduler, while improving the performance of real-world DAG jobs by as much as $3.1\times$ at larger scale.
\end{abstract}

\pdfoutput=1

\section{Introduction}
\label{sec:intro}

In recent years, a new cloud computing model called serverless
computing~\cite{aws_serverless, berkeley_serverless}
(Function-as-a-Service or FaaS)\footnote{
We use the term ``serverless computing'' and ``FaaS'' interchangeably.
} has become prevalent, owing to
OS-level (i.e., container-based) virtualization. 
Serverless computing
enables a new way of building and scaling applications and
services by allowing developers to break traditionally monolithic server-based
applications into finer-grained cloud functions\footnote{
Without loss of generality, we use ``Lambda functions'' to represent cloud functions throughout.
}; developers can thus focus
on developing function logic without having to worry about
provisioning, scaling, and managing traditional backend servers or
VMs, which are notoriously tedious to maintain~\cite{gray_stop}.

With their growth in popularity, serverless computing solutions have found their way into both commercial clouds (e.g., AWS
Lambda, Google Cloud Functions,  
and IBM Cloud
Functions) and open source projects (e.g., OpenLambda~\cite{openlambda_hotcloud16}). 
While serverless platforms were originally intended for event-driven, stateless 
applications~\cite{serverless_usecases},
recent trend has demonstrated the usage of serverless computing 
in support of more complex applications.  

One such example is DAG (directed acyclic graph) workflow-based data analytics applications.
These applications are characterized by short, fine-grained tasks with large 
fan-outs~\cite{sparrow_sosp13, alibaba_trace18, alitrace_iwqos19}. For example, an analysis of Alibaba workload traces shows that more than $50\%$ of the analytics batch jobs, tasks, and instances
are finished within 10 seconds~\cite{alitrace_iwqos19, alitrace_apsys18}.
The auto-scaling property of serverless platforms makes these platforms well-suited for the short, fine-grained tasks and bursty, large fan-outs that characterize DAG-based workflows. In addition, FaaS providers charge users at a fine granularity -- 
AWS Lambda
bills on a per-invocation basis (\$$0.02$ per 1 million invocations)
and charges resource usage by rounding up the function's execution time to the nearest
$100$ milliseconds (ms). 
Workloads with short tasks can take advantage of this fine-grained pay-as-you-go pricing model to keep 
monetary costs low\footnote{Pay-as-you-go in the context of serverless computing is essentially not 
paying for what are not being used.}. 
Consequently, serverless computing can be leveraged as a promising solution for 
next-generation large-scale DAG workloads in high-performance computing (HPC), 
data analytics, and data sciences. 

Moving DAG scheduling from a traditional serverful deployment to the emerging serverless
platforms presents unique opportunities. In a traditional serverful deployment, 
the best practice is to utilize a logically centralized scheduler for managing 
task assignments and resource allocation under various objectives including load balancing,
cluster utilization, fairness and so on. 
State-of-the-art serverful workflow schedulers include but are not limited to: 
MapReduce job scheduler~\cite{mapreduce_osdi04}, 
Apache Spark scheduler~\cite{spark_nsdi12},  Sparrow~\cite{sparrow_sosp13},
and Dask~\cite{dask}. 
In the context of serverless computing, however, the assumptions of the traditional
serverful schedulers do not hold any more. This is because: 
(1) FaaS providers are responsible for managing the ``servers'' (i.e., where the 
task executors are hosted);
and (2) serverless platforms typically provide a \emph{nearly unbounded} amount of
\emph{ephemeral} resources.
As a result, a hypothetical serverless DAG scheduler may not necessarily care about
traditional ``scheduling''-related metrics and constraints (such as load balancing
and cluster utilization), as an individual task could be executed anywhere in the
serverless data center that is essentially managed by the service provider.

Yet, designing a fast and efficient serverless-oriented DAG engine introduces
challenges. First, a task needs to be \emph{dispatched} to a Lambda function as fast as 
possible. With this in mind, a logically centralized scheduler would inevitably introduce 
a performance bottleneck, especially for short-task dominated workloads. 
Second, as already mentioned, serverless platforms come with inherent constraints including
limited outbound-only network connectivity;
as such, a workflow has to rely on an external storage system for storing
intermediate data, which impacts data locality and incurs extra network communications.
Researchers have developed solutions on serverless computing platforms for supporting parallel jobs~\cite{pywren_socc17, excamera_nsdi17, numpywren}; however, these attempts do not fully investigate decentralized DAG scheduling for serverless computing.
Therefore, current state-of-the-art demands a new serverless-native DAG framework optimized to minimize the 
network communication overhead while maximizing data locality whenever possible.

In this paper, we argue that a serverless DAG engine urgently demands a radical 
redesign, with a focus shifted away from the techniques optimized for traditional DAG
schedulers targeting serverful deployments.
To this end, we present {\proj}, a serverless-oriented, 
decentralized, data-locality aware DAG engine. 
{\proj} uniquely exploits the elasticity of the serverless platform (in our case AWS Lambda) and completely delegates the requirements of load balancing, fairness, and resource efficiency to the serverless platform. {\proj} is novel in that it realizes \emph{decentralized scheduling} where a DAG is partitioned into sub-graphs
that are distributed to separate task executors (implemented as an AWS Lambda runtime). 
Lambda runtimes
schedule the tasks in their sub-graphs and cooperate (at ``joins''  
on sub-graph boundaries) to dynamically 
schedule tasks that are in two or more sub-graphs but that must only be executed once. {\proj}
also provides efficient storage mechanisms for managing intermediate data;
however, according to our factor analysis in \cref{subsec:factor}, \proj's decentralized design has the most influence on its performance.
The decentralized design minimizes communication overhead and improves scalability by increasing data locality.

Specifically, we make the following contributions:
\vspace{-3pt}
\begin{itemize}
    \item We thoroughly investigate the problem space of 
    DAG scheduling in serverless computing,
    \item We identify and evaluate a set of techniques to make DAG
    scheduling serverless-aware,
    \item We design and implement {\proj}, a serverless DAG engine attuned to AWS Lambda, 
    \item We evaluate {\proj} to validate its efficacy and design tradeoffs.
\end{itemize}

\pdfoutput=1

\section{Background and Related Work} 
\label{sec:related}

\vspace{-5pt}
\subsection{Serverless Computing Primer}
\label{subsec:primer}
\vspace{-4pt}

\myparagraph{Serverless Computing}
handles virtually all the system administration operations to make it easier
for users and developers to use a near-infinite amount of cloud resources including 
bundled CPUs and memory, object stores, and a lot more~\cite{serverless_berkeley}. 
Service providers provide a flexible function interface so that developers 
can completely focus on development of the core application logic; service providers in turn
help automatically scale the function executions in a demand-driven fashion, hiding the
tedious cluster configuration and management overheads from the users.

\myparagraph{General Constraints and Limitations:}
Service providers place limits on the use of cloud resources to simplify resource management. Take AWS Lambda for example: users have the flexibility of configuring Lambda's memory and CPU resources in a bundle. 
Users can choose a memory amount between 128MB and 3008MB in 64MB increments. 
Lambda allocates CPU power linearly in proportion to the amount of memory configured. Each Lambda function can run at most 900 seconds and will be forcibly returnedwhen function timeouts. In addition, Lambda only allows outbound TCP network connections and bans inbound connections and UDP protocol. 

In addition to these constraints, serverless computing
suffers from a ``cold start'' 
penalty~\cite{peeking_atc18, sock_atc18, sand_atc18}\footnote{
``Cold start'' refers to the first-ever invocation of a function instance.
} associated with container startups. Service providers rely on container caching (i.e., warmed functions) 
to mitigate the impact of cold starts on elasticity. Another limitation that plagues the runtime performance
of serverless applications is the lack of a quality-of-service (QoS) control. As a result,
functions suffer from the straggler issues~\cite{locus_nsdi19}.
\emph{Therefore, an ideal serverless DAG framework should be able to provide effective workaround solutions.}

\myparagraph{Opportunities:}
Running DAG parallel jobs (e.g., distributed linear algebra, 
distributed data analytics, etc.) has long been challenging for domain scientists and 
data analysts due to accessibility, configuration, provisioning, and cluster management 
complexity. The emerging serverless computing model seems to provide an attractive-enough
foundation for potentially relieving the domain scientists and data analysts of the tedious
cluster administration efforts. 
\emph{However, to bridge the gap, current state-of-the-art badly requires a fast and efficient serverless-aware DAG framework middleware.}

\begin{figure*}
\begin{minipage}{\textwidth}
\begin{minipage}[b]{0.31\textwidth}
\begin{center}
\includegraphics[width=1\textwidth]{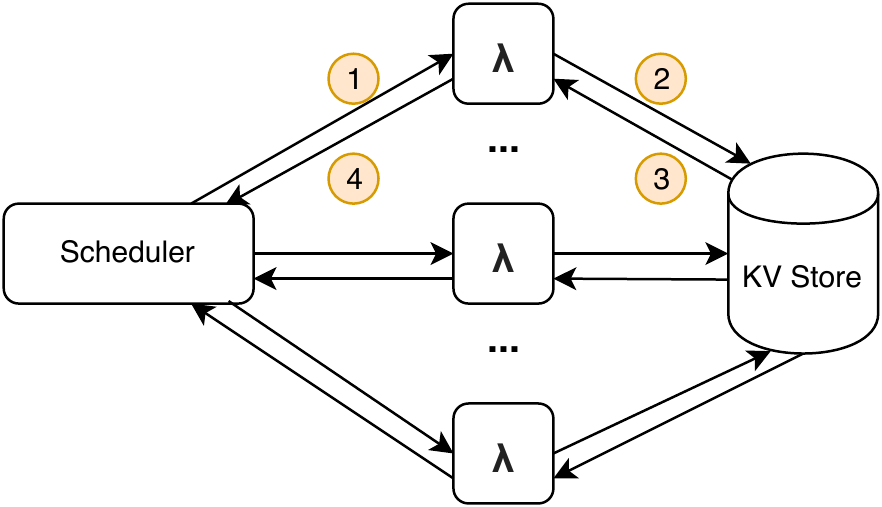}
\end{center}
\vspace{-5pt}
\caption{
The strawman scheduler architecture. Step~1: The scheduler invokes a Lambda 
function, which establishes a TCP connection with the scheduler, executes the task, 
and then in Step~2 sends the output results to the KV store. Once the Lambda function
receives an {\small\texttt{ACK}} from the KV store in Step~3, it notifies the scheduler
in Step~4 that the task is finished.}
\vspace{-10pt}
\label{fig:strawman}
\end{minipage}
\hfill
\hspace{-3em}%
\begin{minipage}[b]{0.3\textwidth}
\begin{center}
\includegraphics[width=1\textwidth]{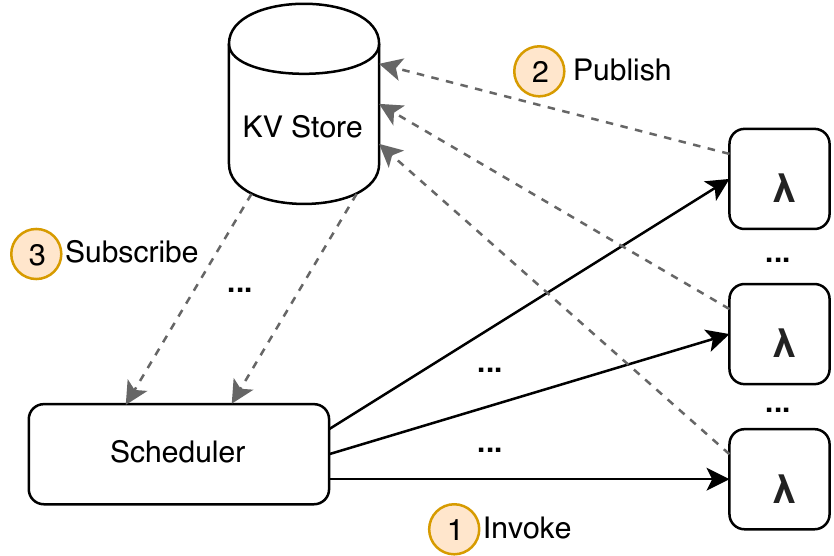}
\end{center}
\vspace{-5pt}
\caption{ 
The pub/sub architecture. In Step~1 The scheduler invokes a Lambda 
function, which executes the task,  and then publishes the output results to the KV store in Step~2. 
The scheduler as the subscriber listens for messages on 
predefined channels, and gets notified in Step~2 whenever a Lambda function publishes the results to the KV store.}
\label{fig:pubsub}
\vspace{-10pt}
\end{minipage}
\hfill
\hspace{-3em}%
\begin{minipage}[b]{0.3\textwidth}
\begin{center}
\includegraphics[width=1\textwidth]{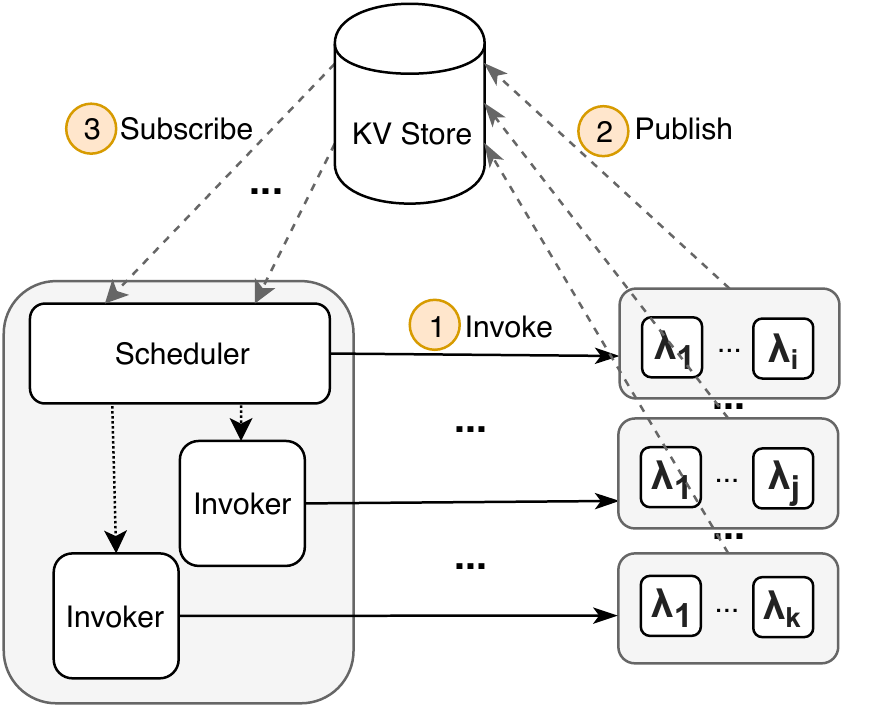}
\end{center}
\vspace{-5pt}
\caption{
The parallel-invoker architecture. This architecture extends the pub/sub architecture in  Figure~\ref{fig:pubsub} with a parallel-invoker that accelerates Lambda invocations by spawning multiple invoker processes in the scheduler to concurrently invoke Lambda functions. }
\label{fig:parallel}
\vspace{-10pt}
\end{minipage}
\end{minipage}
\end{figure*}

\vspace{-4pt}
\subsection{Serverless Workflow Management Frameworks}
\label{subsec:serverless}
\vspace{-3pt}

Existing serverless frameworks have been built using two main approaches. The first is a queue-based master-worker approach in which the master orchestrates the workflow and submits tasks that are ready for execution to a queue. Workers are cloud functions that process these tasks in parallel when possible. For example, numpywren~\cite{numpywren} is a serverless linear algebra framework. numpywren workers are implemented using PyWren ~\cite{pywren_socc17}, which is a framework for executing data-intensive batch processing workloads.

In the second approach, the master directly invokes cloud functions to process ready tasks \cite{HyperFlow, step_functions}. Examples of this include Sprocket~\cite{sprocket_socc18} and ExCamera~\cite{excamera_nsdi17}, which have been developed for serverless video processing. This second approach is also used by general purpose serverless orchestration frameworks. Example frameworks include AWS Step Functions, Azure Durable Functions, Fission Workflows~\cite{fission_workflows}, and the framework in \cite{malawski_serverless_hyperflow_2017}.

L\`{o}pez et~al.~\cite{lopez_comparison_2018} evaluated AWS Step Functions and Azure Durable Functions with respect to their support for parallel execution, among other attributes. They found that overhead for parallelism grows exponentially with the number of parallel functions for AWS Step Functions and Azure Durable Functions.

Fission Workflows is built on top of the Fission~\cite{fission} serverless framework for Kubernetes. 
Users define a DAG by creating a configuration file, which defines tasks and their dependencies. The framework in~\cite{malawski_serverless_hyperflow_2017} is built upon the HyperFlow \cite{HyperFlow} workflow engine. HyperFlow models its workflows using user-written JSON files, and thus is similar to Fission Workflows with respect to how workflows are represented. While manually composing a DAG configuration may work well for coarse-grained microservice-based workflow applications, manually implementing a complex, fine-grained workflow is nontrivial. 
For this reason, \cite{fission_workflows, malawski_serverless_hyperflow_2017}
are not well-suited for supporting complex computing jobs implemented using high-level programming languages.

\proj uses neither a master-worker-queue approach nor a direct invocation approach. Instead, {\proj} adopts a decentralized approach in which the global DAG is partitioned into local subgraphs. Each \proj executor is responsible for scheduling and executing the tasks within its assigned subgraph in an autonomous manner. An executor assumes the master's role when it uses its assigned subgraph to determine when its tasks can be executed; an executor assumes a worker's role when it executes these tasks. \proj executors coordinate to ensure that the dependencies in the global DAG are satisfied.

\pdfoutput=1

\vspace{-6pt}
\section{Motivational Study: A Journey from the Serverful to the Serverless}
\label{sec:moti}
\vspace{-4pt}

Prior to the emergence of the serverless
computing model, DAG schedulers were designed to work with a finite number of
compute and storage resources.
These schedulers have to maintain a 
(complete or partial) global view of which tasks are running where, and use this 
view to optimize with respect to certain predefined objectives. 
Serverless computing, on the
other hand, offers a nearly infinite amount of ephemeral resources, which are
transparently managed by the service provider. Consequently, traditional schedulers 
would fail to utilize cloud resources optimally.
{\proj} takes a radically different approach and is motivated by the above
observations with the goal of improving the performance for task dispatching with 
respect to serverless platforms. In this section, we present our motivational study
of designing a fast and efficient serverless DAG engine.

\vspace{-5pt}
\subsection{A Strawman Scheduler}
\label{subsec:strawman}
\vspace{-3pt}

We began our journey by implementing 
a centralized DAG scheduler, which simply parsed
the user-defined job code, generated a DAG data structure, and sent off the DAG tasks to a group of Lambda functions for execution. 
Our strawman scheduler was a modification of the Python-written Dask
{\small\texttt{distributed}} scheduler. Dask is an open-source parallel computing library for Python data analytics~\cite{dask}. Traditionally, Dask 
{\small\texttt{distributed}} executes tasks within so-called worker processes, each 
running as a long-lived server across a cluster of machines. In Dask,
the scheduler sends tasks to the worker processes for execution.
Worker processes run as long-lived servers across a cluster of
machines. The Dask {\small\texttt{distributed}} scheduler uses a communication
protocol to communicate with workers and balance their load with
respect to certain optimization constraints, such as data locality
and memory consumption. We reused the 
DAG and communication protocol modules from Dask, used AWS
Lambda for task execution instead of worker processes, and disabled load balancing as load balancing is handled by AWS Lambda. 

In a typical serverful distributed processing framework, worker processes can
directly communicate with each other using TCP. A worker process that needs to execute a task {\small\texttt{T}} may find that {\small\texttt{T}}'s input data is not stored locally. This worker will then issue TCP requests for {\small\texttt{T}}'s input data to the workers who executed the upstream tasks that output this data.
In our serverless computing environment, Lambda functions are not allowed to accept inbound TCP connection requests. Due to this constraint,
the upstream tasks in \proj would have to store their output data in external distributed storage
(a key-value store or KV store in short),
from which the dependent downstream tasks can read their input data and make progress.
Figure~\ref{fig:strawman} depicts the strawman approach.

\vspace{-5pt}
\subsection{Publish/Subscribe Model}
\label{subsec:pubsub}
\vspace{-3pt}

While the centralized strawman scheduler worked, it suffered from several performance bottlenecks.
The first performance bottleneck was due to the large number of concurrent TCP connection requests sent to the scheduler from the Lambda functions.
A short-lived Lambda function will  immediately request a TCP connection with the scheduler to acknowledge the completion of its task. This makes it easy for a pool of thousands of newly invoked Lambda functions to overwhelm the scheduler. This is not a problem for a serverful deployment, e.g., a statically deployed Hadoop cluster with hundreds of worker nodes that 
established TCP connections at cluster initialization phase.

To address this problem, we adopted a pub/sub (publisher/subscriber) approach (Figure~\ref{fig:pubsub}). The pub/sub scheduler provided higher performance than the strawman scheduler, since
sending task completion messages through pub/sub channels was more efficient than using a
large number of concurrent TCP connections; also, the number of network hops was reduced. The pub/sub architecture was easy to integrate, since external storage was already being used to store intermediate results.

\vspace{-5pt}
\subsection{+Parallel Invokers}
\label{subsec:parallel}
\vspace{-3pt}

While the pub/sub approach had substantially improved network performance, the framework struggled to launch Lambda functions quickly enough for large, bursty workloads.
This is due to the large cost of invoking a Lambda function
(e.g., invoking an AWS Lambda function takes about 50 milliseconds with the Boto3 AWS Python API). 
To scale-up  Lambda invocation performance, we created a large number of dedicated Lambda-invoker processes co-located with the scheduler (Figure~\ref{fig:parallel}).
When DAG task dependencies resolve, the scheduler evenly distributes task invocation responsibilities 
among multiple invoker processes, enabling (near-)linear speedup.

\begin{figure}[t]
\begin{center}
\includegraphics[width=0.45\textwidth]{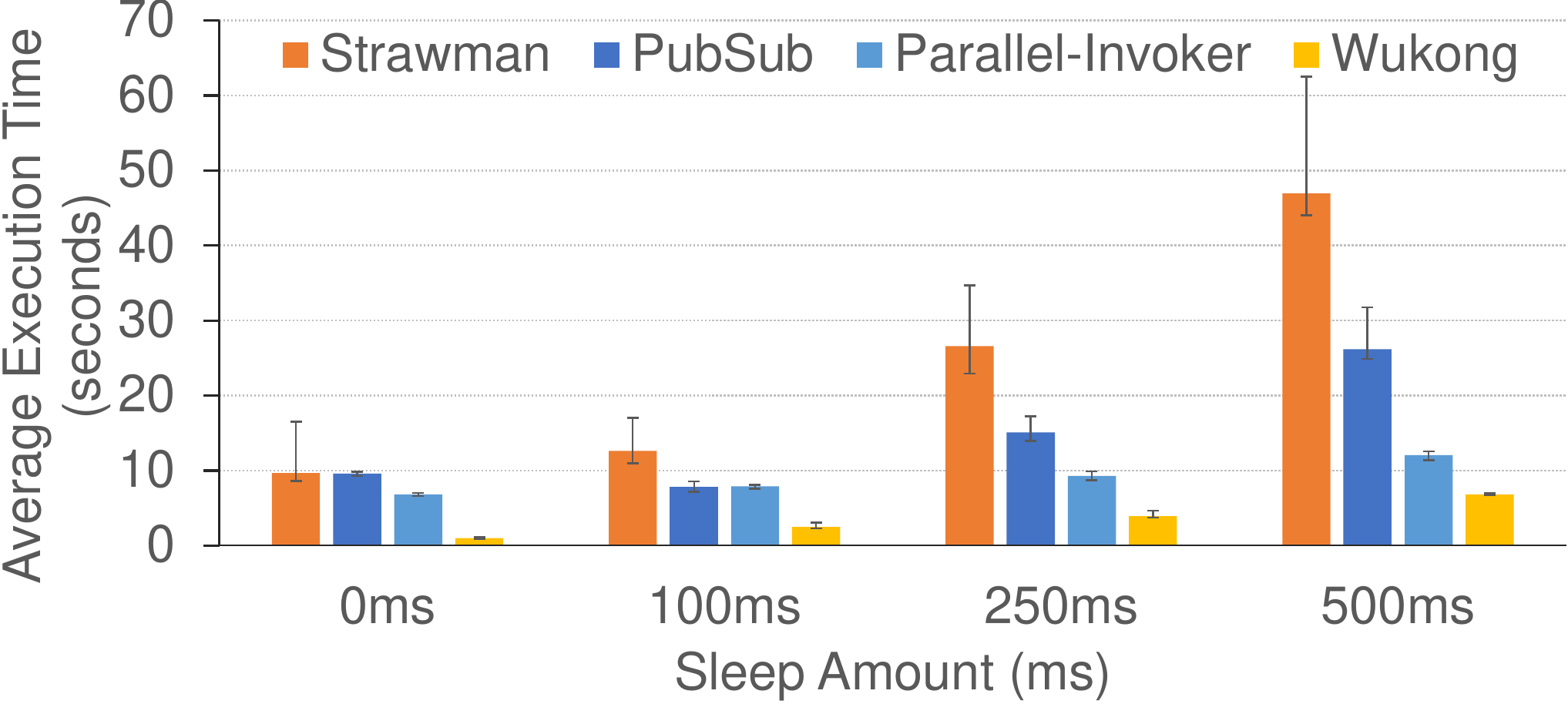}
\label{TreeReduction}
\vspace{-5pt}
\caption{
Performance comparison of different design iterations for Tree Reduction (TR). TR is a microbenchmark with a tree-like DAG topology~\cite{dask_tr}, which combines neighboring elements until there is only one left. We ran TR with an initial array of 1024 numbers (i.e., 512 leaf tasks at the bottom of the DAG) on each system ten times and recorded the average (bars), and \{min,  max\} (error bars). We intentionally added sleep-based delays in each task to simulate a compute task with a controllable duration. 
}
\vspace{-25pt}
\label{fig:moti}
\end{center}
\end{figure}

Figure~\ref{fig:moti} plots the average execution time achieved by each design iteration. 
We intentionally added a sleep-based delay to each task in order to simulate a compute task with controllable duration.
For TR test with 0ms sleep delays, the performance difference between the strawman and pub/sub versions of our framework is roughly the same due to the fact that the TR workload is primarily dominated by the communication overhead of transferring the array over the network. 
As noted above, the parallel-invoker version is able to execute TR $24.2\%$ faster than strawman and pub/sub. This is because TR is also characterized by a large number of leaf tasks. Specifically, the TR algorithm generates \(\frac{n}{2}\) leaf tasks, where \(n\) is the length of the input array. The parallel-invoker version can invoke the leaf tasks at a significantly higher rate than strawman and pub/sub; consequently, parallel-invoker performs better for workloads with a large number of leaf tasks. 
As the time span of each task increases, pub/sub start to show performance benefit against strawman, because a less number of TCP connections significantly reduces the amounts of IRQ requests which flood the strawman case. 
Parallel-invoker improves the performance against pub/sub, but is still sub-optimal due to network I/O overheads. The goal of {\proj} is to reduce the execution time of a DAG job, an optimal serverless DAG engine that dispatches DAG tasks with minimum runtime overhead.

One critical issue of the parallel-invoker architecture is its large commitment of resources
to the centralized pub/sub scheduler throughout the whole course of the workload. 
Due to this, we moved to a decentralized scheduler design.  This major design change came about as a result of a key observation, which was that the scheduler was  only being used to launch 
downstream tasks as data dependencies of the DAG were resolved. Instead of scaling-up the invocation process of the centralized scheduler, each Lambda function could directly handle the  responsibility of invoking  downstream tasks without having to coordinate with the centralized scheduler. This lead to an effective, serverless-aware, scale-out design that is described next.

\pdfoutput=1

\vspace{-5pt}
\section{{\proj} Design}
\label{sec:design}
\vspace{-3pt}

In this section, we present the system design of {\proj}. We describe the high-level components and discuss the techniques used for static scheduling, task execution, dynamic scheduling, and managing storage.

\begin{figure}[t]
\begin{center}
\includegraphics[width=0.45\textwidth]{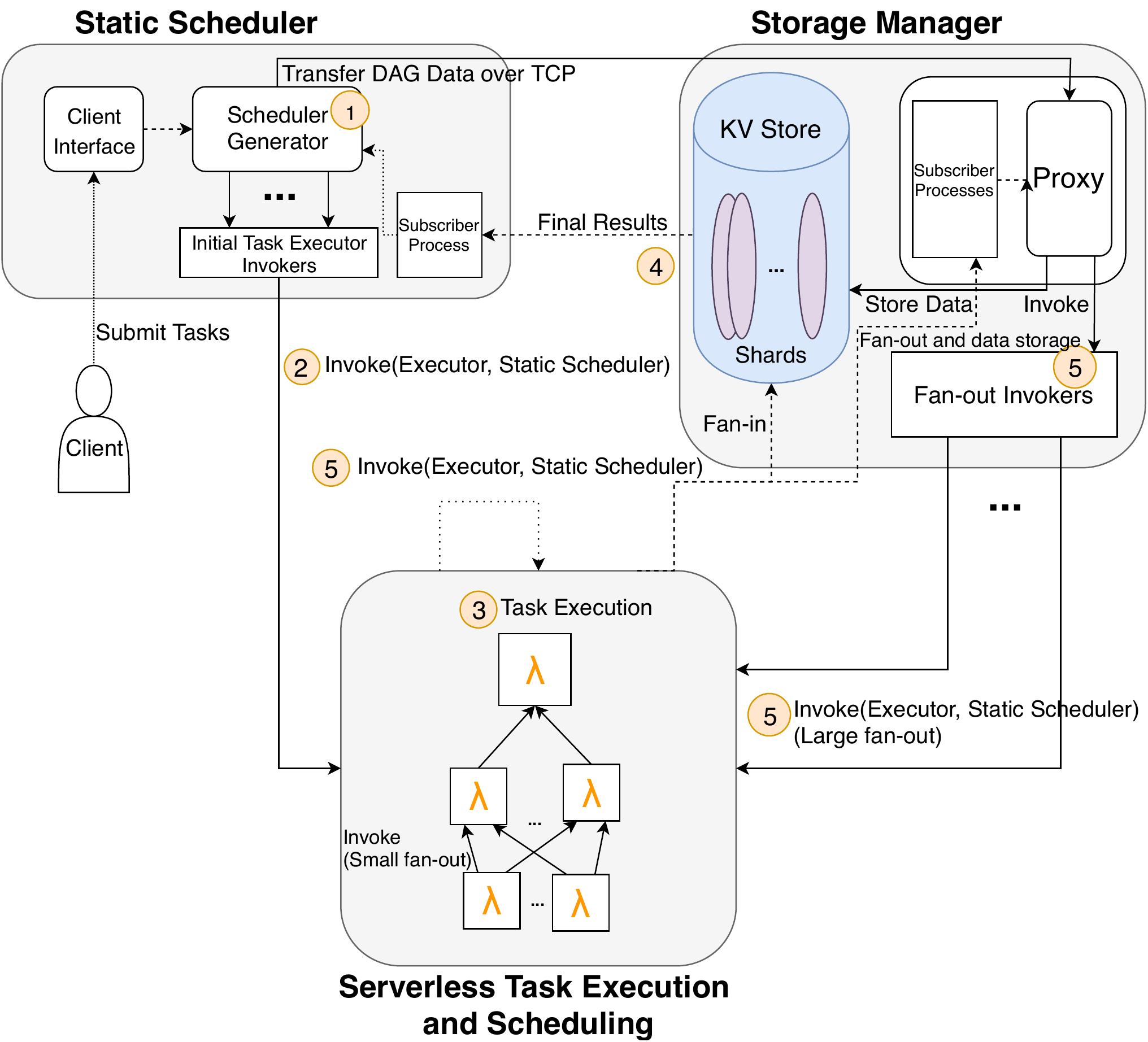}
\vspace{-5pt}
\caption{Overview of {\proj} architecture.}
\vspace{-20pt}
\label{fig:arch}
\end{center}
\end{figure}

\subsection{High-Level Design}
\label{subsec:HighLevelDesignAndExecution}

{\proj} consists of three major components: a static scheduler, a serverless task execution and scheduling runtime, and a storage manager. Figure~\ref{fig:arch} shows the high-level design of {\proj}. This figure reflects a major design revision to the Pub/Sub scheduler described at the end of \cref{sec:moti}. This revision removed the requirement for Lambda functions to acquire downstream tasks from the KV store.
We modified the scheduler to produce static schedules, where each schedule represents a sub-graph of the DAG. A static schedule contains all the task code and other required (static) information, e.g., data dependencies, for each task in the sub-graph. The scheduler now passes a static schedule to each Lambda function it invokes, meaning that each function starts with all of the task code that it may have to execute. This removed the necessity for Lambda functions to grab downstream task code from the KV store,
which speeds up execution by decentralizing {\proj}. Since static schedules contain the dependencies, scheduling operations for fan-in and fan-out processing can be done dynamically by the Lambda functions, which removes the need for a centralized scheduler to determine when data dependencies have been satisfied.

\subsection{Static Scheduling}
\label{subsec:StaticScheduling}

\begin{figure}[t]
\begin{center}
\includegraphics[width=0.45\textwidth]{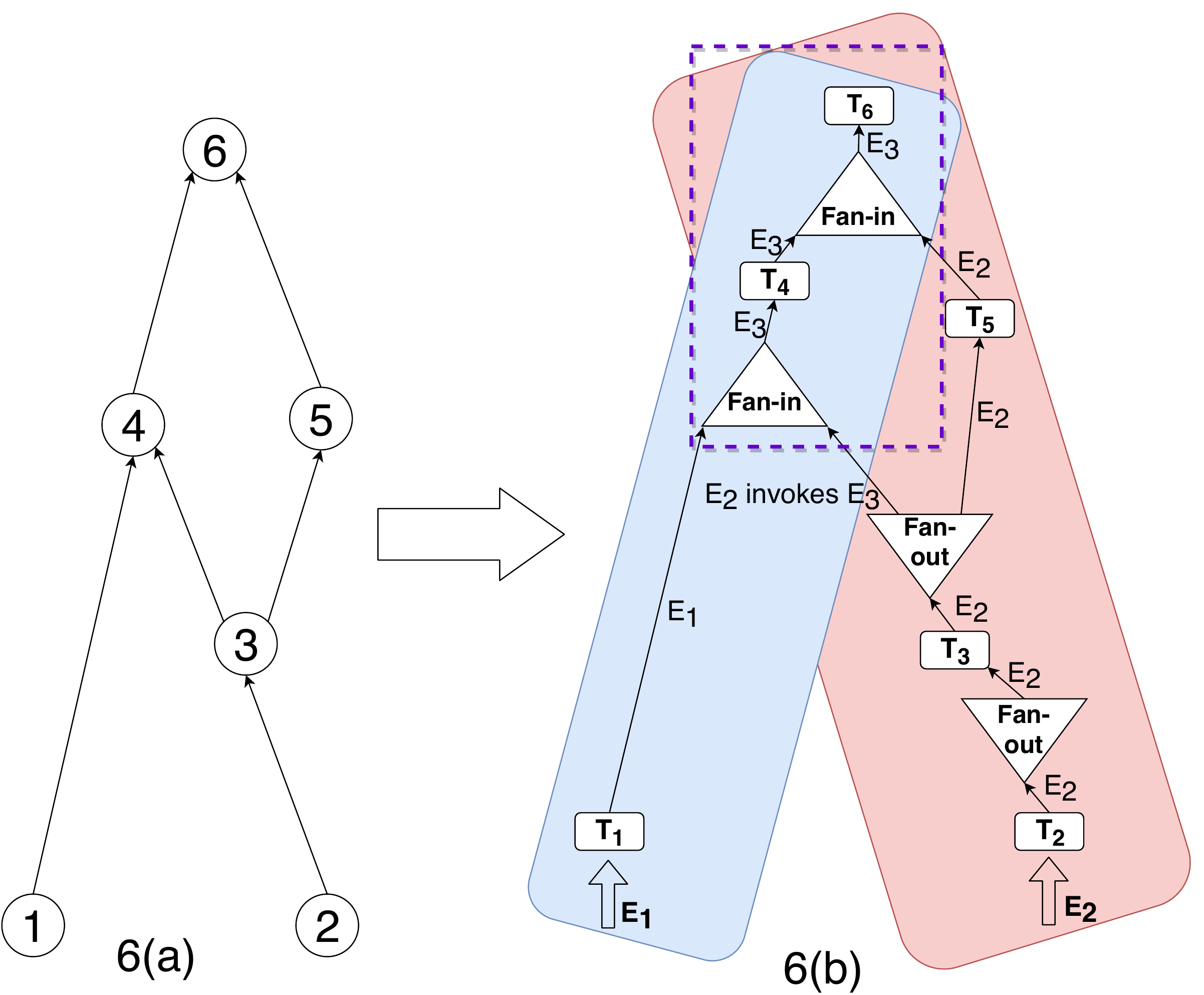}
\vspace{-5pt}
\caption{Static and dynamic scheduling.
}
\vspace{-20pt}
\label{fig:sched}
\end{center}
\end{figure}

{\proj} users submit a Python computing job to {\proj}'s DAG generator, which converts the job into a DAG. The static Schedule Generator generates static schedules from the DAG. For a DAG with $n$ leaf nodes, $n$ static schedules are generated. A static schedule for leaf node {\small\texttt{L}} contains all of the task nodes that are reachable from {\small\texttt{L}} and all of the edges into and out of these nodes. The data for a task node includes the task's code and the KV Store keys for the task's input data. The schedule for {\small\texttt{L}} is easily computed using a depth-first search (DFS) that starts at {\small\texttt{L}}. Figure~\ref{fig:sched}(a) shows a DAG with two leaf nodes. Figure~\ref{fig:sched}(b) shows the two static schedules that are generated from the DAG --- {\small\texttt{Schedule 1}} is the nodes and edges in the region colored blue (left) and {\small\texttt{Schedule 2}} is the nodes and edges in the region colored red (right). 

Static schedules are used to reduce the number of network I/O operations that must be performed by the Task Executors, which improves overall system performance. Instead of executing a single task and retrieving the next task from the KV store, Task Executors receive a static schedule of all of the tasks (including the task code) they may possibly execute. 

A static schedule contains three types of operations: task execution, fan-in, and fan-out. Note that there is at least one fan-in or fan-out operation between each pair of tasks. To simplify our description, when tasks {\small\texttt{T1}} is followed immediately by task {\small\texttt{T2}} in a DAG and {\small\texttt{T1}} ({\small\texttt{T2}}) has no fan-out (fan-in), we add a trivial fan-out operation between {\small\texttt{T1}} and {\small\texttt{T2}} in the static schedule. This fan-out operation has one incoming edge from {\small\texttt{T1}} and one outgoing edge to {\small\texttt{T2}}, i.e., there is no actual fan-out. In Figure~\ref{fig:sched}(a) and Figure~\ref{fig:sched}(b), this is the case for tasks {\small\texttt{T2}} and {\small\texttt{T3}}.

A task operation may appear in more than one static schedule. In Figure~\ref{fig:sched}(b), tasks {\small\texttt{T4}} and {\small\texttt{T6}} are both in {\small\texttt{Schedule 1}} and {\small\texttt{Schedule 2}}. This will create a scheduling conflict between the Task Executors that are assigned to schedule these tasks, since tasks {\small\texttt{T4}} and {\small\texttt{T6}} should be executed  only once. There is not enough information available in the DAG to statically determine how to resolve scheduling conflicts so that execution time is minimized; instead, conflicts are resolved by dynamic scheduling operations performed by the Task Executors. Note also that a static schedule does not map a given task {\small\texttt{T}} to a processor;
this mapping is done dynamically and automatically by the AWS Lambda runtime when the Task Executor that will (eventually) execute task {\small\texttt{T}} is invoked. A static schedule only specifies a valid partial-ordering of the tasks it contains --- tasks are to be executed in bottom-up order, starting with the leaf node in the static schedule. Dynamic scheduling during task execution imposes the remaining constraints on task order. The time and place that tasks are executed is determined at runtime.

\subsection{Task Execution and Dynamic Scheduling}
\label{subsec:DynamicScheduling}

Execution starts when the scheduler's Initial Task Executor Invokers assign each static schedule produced by the Schedule Generator to a separate AWS Lambda function, which we refer to as a \emph{Task Executor}, and invoke the set of initial Task Executors in-parallel. Each of these Task Executors performs the first operation in its static schedule, which is always to execute the leaf node task in its static schedule. In Figure~\ref{fig:sched}(b), Task executors {\small\texttt{E1}} and {\small\texttt{E2}} execute leaf tasks {\small\texttt{T1}} and {\small\texttt{T2}}, respectively. An Executor will then execute the tasks along a single path in its static schedule, enforcing the static ordering of tasks along the path. 

If Task Executor {\small\texttt{E}} executes a fan-out operation with only one out edge, the operation has no effect --- Executor {\small\texttt{E}} simply performs the next operation in its schedule, which will be to execute the next task. A Task Executor may thus execute a sequence of tasks before it reaches a fan-out operation with more than one out edge or a fan-in operation. For such a sequence of tasks, there is no communication required for making the output of the earlier tasks available to later tasks for input. All intermediate task outputs are cached in the local memory of the the Task Executor.

If Task Executor {\small\texttt{E}} executes a fan-out operation with $n$ (where $n > 1$) out edges, then {\small\texttt{E}} invokes $n-1$ new Task Executors. The intermediate output objects that are needed by the new Executors are sent to the Storage Manager for storage in the KV Store, and the associated keys are passed to the invoked Executors as arguments. Each of the $n-1$ Executors will be assigned a static schedule that begins with one of the $n-1$ out edges. Each of these (possibly overlapping) static schedules corresponds to a sub-graph of {\small\texttt{E}}'s static schedule. Executor {\small\texttt{E}} continues task execution and scheduling along the remaining out edge and executes the next operation encountered on this edge. We say that {\small\texttt{E}} becomes the Executor for one out edge and invokes Executors for the remaining $n-1$ out edges. In Figure~\ref{fig:sched}(b), each fan-out operation has one edge labeled ``becomes'' and $0$ or more out edges labeled ``invokes''. The label on an in or out edge also indicates the Executor that is performing the dynamic scheduling operations that involve that edge. For Executor {\small\texttt{E2}}, the first fan-out operation is trivial. On {\small\texttt{E2's}} second fan-out operation, {\small\texttt{E2}} becomes the Executor that will execute {\small\texttt{T5}} and invokes Executor {\small\texttt{E3}}.

As mentioned above, a fan-in operation represents a scheduling conflict between two or more Task Executors that are executing overlapping static schedules. If Task Executor {\small\texttt{E}} executes a fan-in operation with $n$  (where $n > 1$) in-edges, then {\small\texttt{E}} and the $n-1$ other Task Executors involved in this fan-in operation cooperate to see which one of them will continue their static schedules on the out edge of the fan-in. The Task Executors that do not continue will send their intermediate output objects to the Storage Manager and stop. In Figure~\ref{fig:sched}(b), the first fan-in operation of Executors {\small\texttt{E1}} and {\small\texttt{E3}} resolves the conflict between their static schedules. We assume that {\small\texttt{E3's}} fan-in operation is executed after {\small\texttt{E1}}'s fan-in operation; thus, {\small\texttt{E1}} stops and {\small\texttt{E3}} continues executing its static schedule at the out edge of the fan-out. At {\small\texttt{E3's}} next fan-in operation, which also involves {\small\texttt{E2}}, we assume {\small\texttt{E3}} executes its fan-in operation last so that {\small\texttt{E2}} stops and {\small\texttt{E3}} executes task {\small\texttt{T6}} and then stops. 

Task Executors cooperate on fan-in operations for a fan-in {\small\texttt{F}} by accessing an associated dependency counter for {\small\texttt{F}} that is stored in the KV Store. This counter tracks the number of {\small\texttt{F}}'s input dependencies that have been satisfied during execution. When a Task Executor {\small\texttt{E}} finishes the execution of a task that is one of the input dependencies of {\small\texttt{F}}, Executor {\small\texttt{E}} performs an atomic increment operation on the dependency counter of {\small\texttt{F}}. The updated value of the dependency counter is then compared against the number of input dependencies of {\small\texttt{F}}. If the value of the dependency counter is equal to the number of input dependencies, then all input dependencies of {\small\texttt{F}} have been satisfied and the task {\small\texttt{T}} on the out edge of {\small\texttt{F}} is ready for execution. In this case, task {\small\texttt{E}}, which executed the last dependent task of the fan-in, will continue its static schedule by executing {\small\texttt{T}}. If, instead, the value of the dependency counter is less than the number of input  of {\small\texttt{F}}, then some input dependencies of {\small\texttt{F}} have yet to be satisfied. In that case, task {\small\texttt{T}} is not ready for execution, so {\small\texttt{E}} saves its intermediate output objects and stops. Notice that no Task Executor waits for any input dependencies of a fan-in to be satisfied. Note that AWS Lambda would bill Task Executors for wait time, which is why waiting is avoided.

For fault tolerance, we relied on the automatic retry mechanism of AWS Lambda, which allows for up to two automatic retries of failed function executions. In the future, we will investigate more advanced error handling mechanisms.

\vspace{-5pt}
\subsection{Storage Management}
\label{subsec:Storage Management}
\vspace{-3pt}

The Storage Manager performs various storage operations on behalf of the Task Executors and the Scheduler. At the start of workflow processing, the Storage Manager receives the workflow DAG and the static schedules derived from the DAG from the Scheduler.

\myparagraph{Intermediate and Final Result Storage:} Task Executors publish their intermediate and final task output objects to the KV Store. Final outputs are relayed to a Subscriber process in the Scheduler for presentation to the Client.

\myparagraph{Small Fan-out Task Invocations:} When a Task Executor performs a fan-out operation that has a small number of out edges, the Task Executor will make the necessary Executor invocations itself. However, sequentially executing a large number of invocations is time consuming so the Executor Task invocations are performed in parallel with assistance from the Storage Manager.

\myparagraph{Large Fan-out Task Invocations:} When a fan-out has a number of out edges that is larger than a user-specified threshold, the Task Executor publishes a message that is relayed to a Subscriber process in the Storage Manager, which passes the message to the Proxy. This message contains an ID that identifies the fan-out's location in the DAG. The Proxy uses the DAG and the fan-out ID to identify the fan-out's out edges in the DAG. This allows the Proxy, with the assistance of the Fan-out Invokers in the Storage Manager, to make the necessary Task Executor invocations, in parallel. The Proxy passes to each Executor its intermediate inputs (or their key values in the KV Store) and the Executor's static schedule.

\pdfoutput=1

\section{Preliminary Results}
\label{sec:eval}

We have implemented {\proj} using roughly $6,000$ lines of Python code (about $2,500$ LoC for the AWS Lambda Runtime, $2,400$ LoC for the Storage Manager, and $1,100$ LoC for the Static Scheduler). {\proj} currently supports AWS Lambda. Porting {\proj} to other public cloud and open source platforms is our work in progress.

\myparagraph{Experimental Goals and Methodology.}
The goals of our preliminary evaluation were to:
\begin{itemize}
\item identify and describe the main factors influencing the performance and scalability of {\proj},
\item compare {\proj} against the serverful Dask framework to determine whether \proj can achieve comparable performance, even with the inherent limitations imposed by AWS Lambda.
\end{itemize}

We compare the performance of {\proj} against Dask {\small\texttt{distributed}} on two different setups: a five-node EC2 cluster with each virtual machine (VM) running five worker processes and a local setup on a laptop with four worker processes.
We repeated the same tests on an easy-to-use laptop computer to further demonstrate that, with the same workload, {\proj} can achieve superior performance with minimal cluster administration effort.

Our evaluation was performed on AWS. The static scheduler ran in a {\small\texttt{c5.18xlarge}} EC2 VM and the KV Store was a Redis cluster partitioned across ten {\small\texttt{c5.18xlarge}} shards. The KV Store proxy was co-located on the same VM as one of the ten Redis shards. Each Lambda function was allocated 3GB memory with a timeout parameter set to two minutes.

Each node in the five-node cluster was an EC2 {\small\texttt{t2.2xlarge}} VM. We configured this cluster with general-purpose VMs to see if our serverless platform could match their performance. We opted to not configure a cluster of increased price and performance as we cannot configure our AWS Lambda functions to match the processing power of such a cluster. Further, we cannot control for the various restrictions held in place by Amazon including the rate at which we can invoke Lambda functions, the memory allocated to Lambda functions (above 3GB), the network resources allocated to each function, etc. 
The laptop was equipped with a two-core {\small\texttt{Intel i5}} CPU @ \textit{2.30GHz}. Each Dask worker was allocated 2GB of laptop memory.

We describe the tested DAG applications as follows.

\myparagraph{Tree Reduction (TR):} TR sums the elements of an array. TR repeatedly adds adjacent elements until only a single element remains. The implementation used here is general-purpose; it is not optimized for a highly distributed, serverless algorithm, serving as a microbenchmark for effectively evaluating serverless DAG engine. 

\myparagraph{General Matrix Multiplication (GEMM):} GEMM, as the core of many linear algebra algorithms, performs matrix multiplication. We evaluate the performance of \proj for GEMM with two different matrix sizes: \(10,000 \times 10,000\) and \(25,000 \times 25,000\).

\myparagraph{Singular Value Decomposition (SVD):} Two SVD workloads are used. The first workload computes the SVD of a tall-and-skinny matrix, i.e., a matrix with a significantly larger number of rows than columns (SVD1). The second workload computes the rank-5 SVD of an $n \times n$ matrix using an approximation algorithm (SVD2) provided by~\cite{halko_finding_2009}. We use both SVD workloads as a real-world application to evaluate the performance of \proj for increasingly large SVD problem sizes. 

\myparagraph{Support Vector Classification (SVC):} SVC is a real-world machine learning application. We evaluate the performance of \proj on SVC with increasingly large problem sizes. This workload is a benchmark that was retrieved from the publicly available Dask-ML benchmarks~\cite{dask-ml}. 

\vspace{-5pt}
\subsection{End-to-End Performance Comparison}
\label{subsec:e2e}
\vspace{-3pt}

\begin{figure}[t]
\begin{center}
\includegraphics[width=0.4\textwidth]{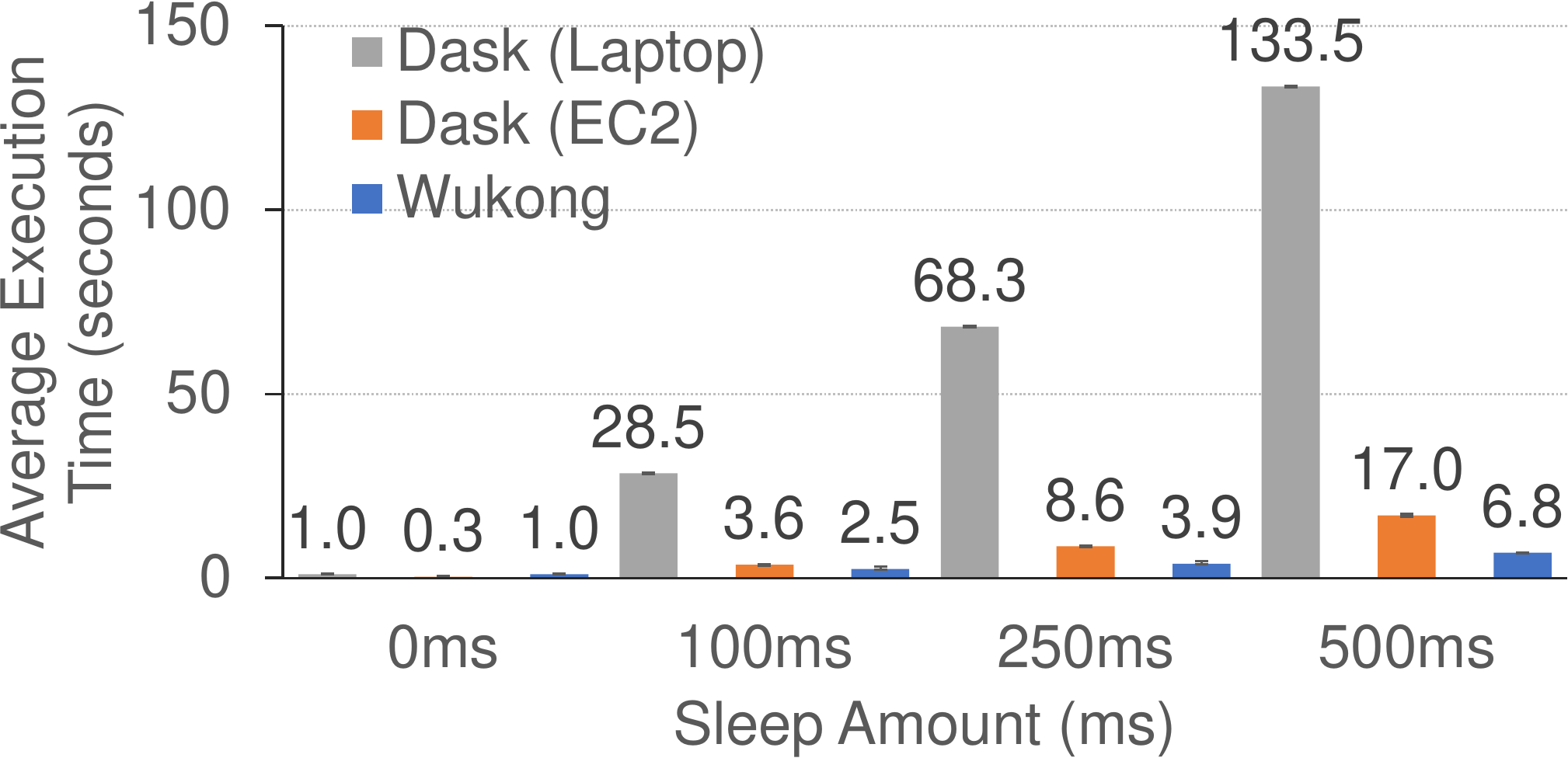}
\vspace{-5pt}
\caption{TR performance comparison.}
\vspace{-20pt}
\label{fig:tr}
\end{center}
\end{figure}

As mentioned earlier, serverless computing suffers from cold starts. We address this issue by warming up a pool of Lambdas, which is the same strategy employed by ExCamera \cite{excamera_nsdi17}. 
Due to AWS' planned cold-start performance optimizations for Lambdas running within a virtual private cloud \cite{munns_announcing_nodate}, ``cold start" penalties should not be nearly as large of an issue in the future.

We first examine the performance of TR (for a preliminary analysis that compares the various design iterations please refer to Figure~\ref{fig:moti} in \cref{subsec:parallel}).
As shown in Figure~\ref{fig:tr}, {\proj} greatly outperforms all previous versions of the framework. The decentralized scheduler reduces the network I/Os required to complete the workload; however, due to the extremely short-duration add operations used by TR with 0ms sleep delays, the communication overhead of transferring the underlying array greatly outweighs the performance gains from increased parallelism. This is why {\proj} achieves lower performance than Dask (EC2).
{\proj} outperforms all other execution platforms when small sleep delays are added to each operation of the tree reduction. {\proj} executes $2.5\times$ faster than Dask (EC2) in the case of 500ms delays. These small delays simulate additional work for each task. The results of this workload with added delays indicate that for workloads with longer tasks, the increased parallelism provided by {\proj} outweighs the communication overhead, demonstrating that our decentralized DAG scheduler incurs minimum overheads.

\begin{figure}[t]
\begin{center}
\includegraphics[width=0.4\textwidth]{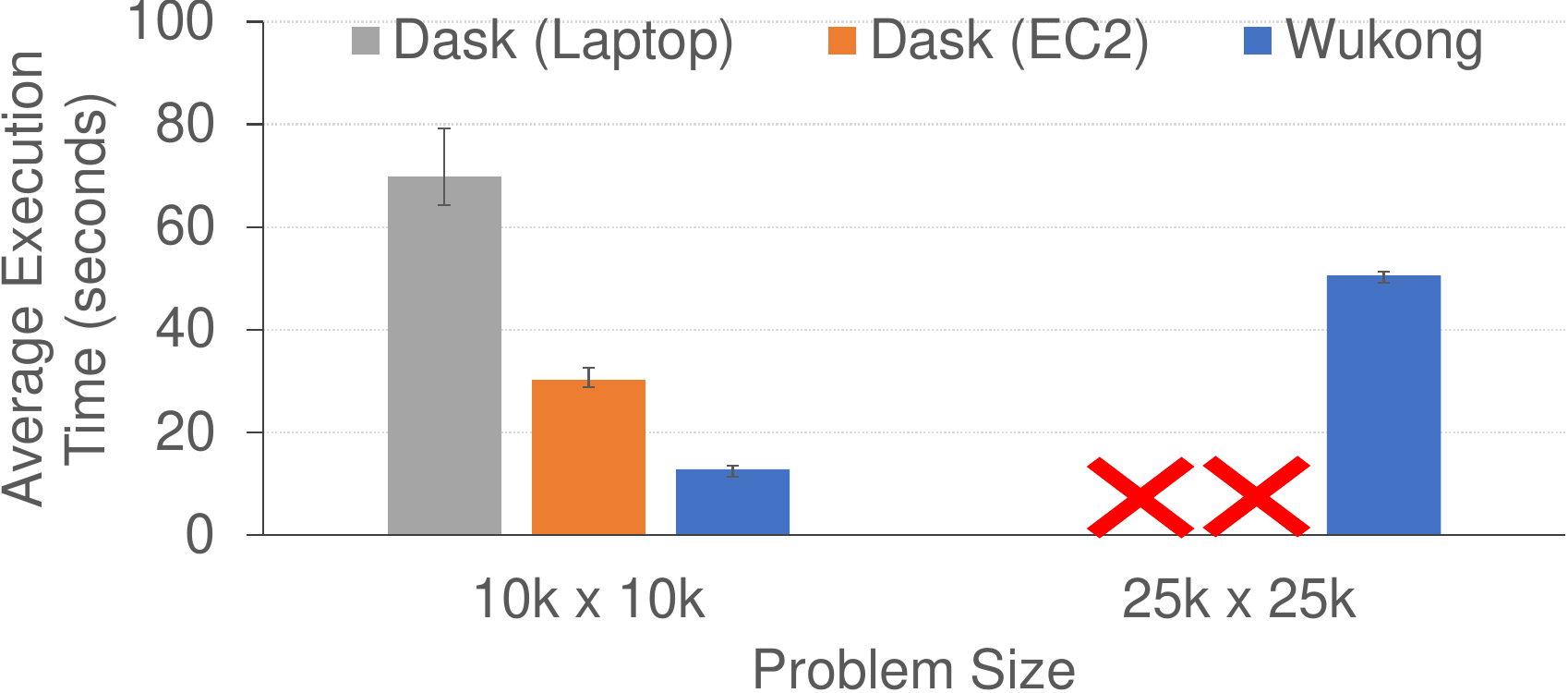}
\vspace{-5pt}
\caption{GEMM performance comparison.}
\vspace{-20pt}
\label{fig:gemm}
\end{center}
\end{figure}

The results of our GEMM tests further demonstrate \proj's superiority in elasticity and performance.  
In the case of case of $10,000 \times 10,000$ matrix multiplication, \proj executed the workload more than twice as fast as Dask (EC2) and more than five times as fast as Dask (Laptop). Dask (EC2) could likely perform this workload faster if the cluster was larger, whereas for {\proj}, it leverages the large number of CPUs provided by AWS Lambda to elastically scale up the performance. When multiplying $50,000 \times 50,000$ matrices, both setups of Dask (Laptop and EC2) suffered from out-of-memory (OOM) errors, failing to complete the job. Our analysis of GEMM on {\proj} indicates that these workloads were dominated by the communication overhead of transferring portions of the matrix to the Task Executors. 

\begin{figure}[t]
\begin{center}
\includegraphics[width=0.4\textwidth]{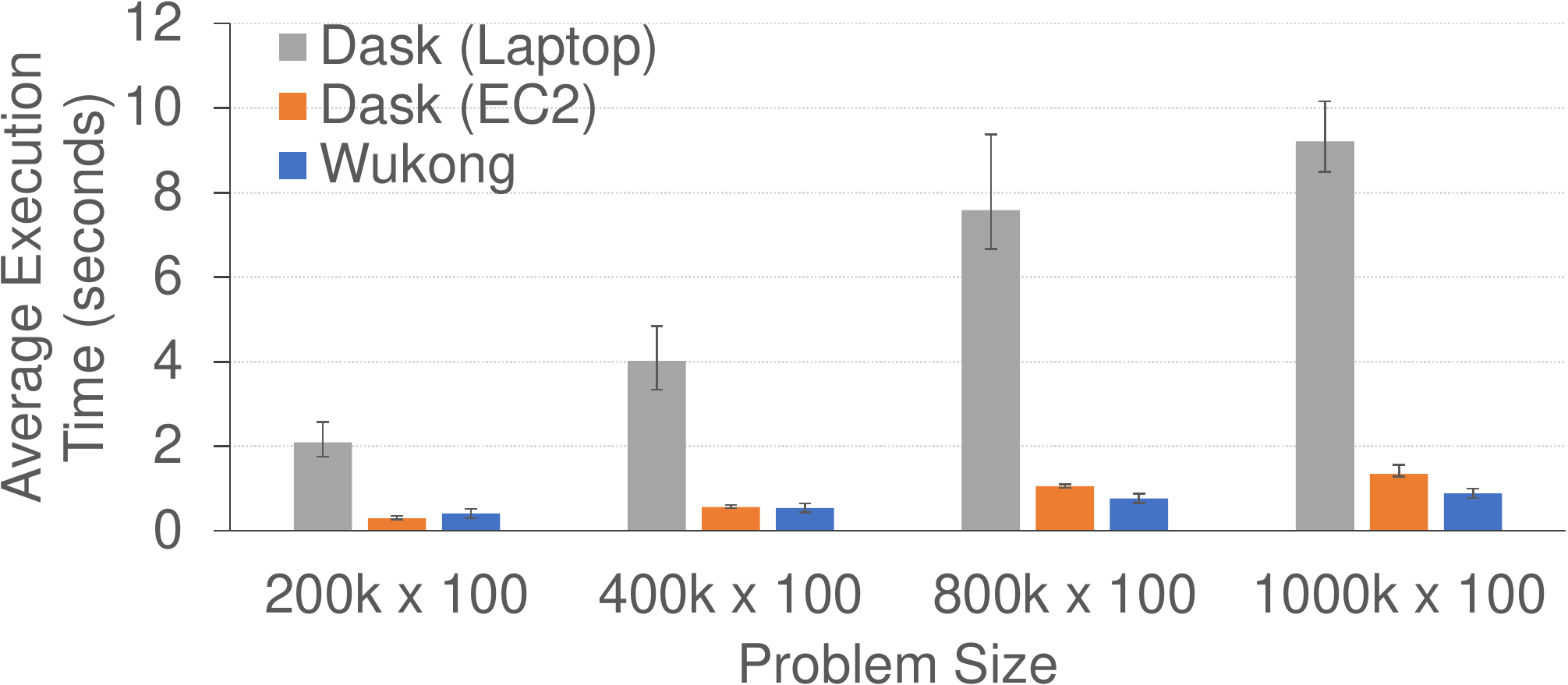}
\vspace{-5pt}
\caption{SVD1: SVD of tall-and-skinny matrix.}
\vspace{-20pt}
\label{fig:svd1}
\end{center}
\end{figure}

\begin{figure}[t]
\begin{center}
\includegraphics[width=0.4\textwidth]{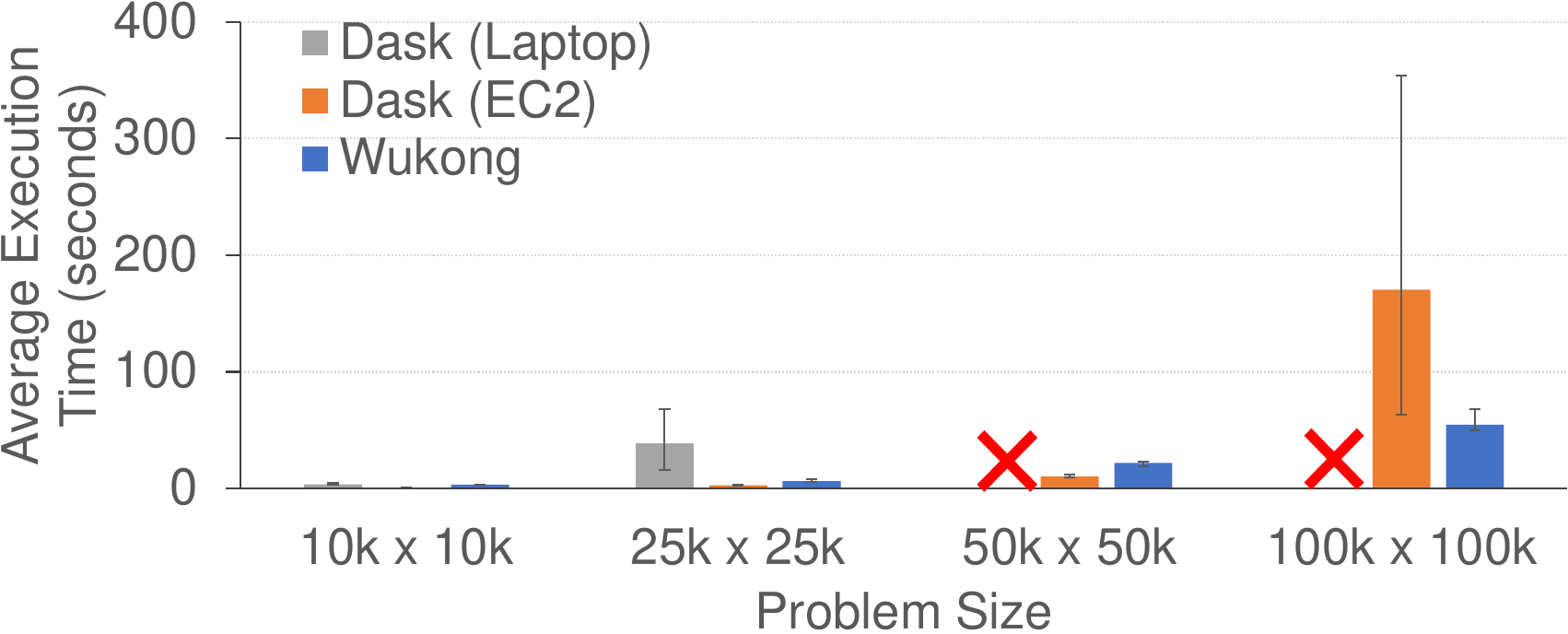}
\vspace{-5pt}
\caption{SVD2: SVD of general matrix.
}
\vspace{-20pt}
\label{fig:svd2}
\end{center}
\end{figure}

Next we analyze the performance of the two SVD workloads. For SVD1, we used the following numbers of rows: $200k$, $400k$, $800k$, and $1,000k$. Figure~\ref{fig:svd1} shows that both Dask (EC2) and \proj were able to greatly outperform Dask (Laptop). For the first two problem sizes, Dask (EC2) out-performed \proj; however, as the problem size increased, the performance of \proj began to exceed that of Dask (EC2). This is because the parallelism from AWS Lambda began to outweigh the communication overhead of the workload. Even so, the overhead associated with network I/Os was a significant factor in the performance of this workload on \proj. 

The dominance of communication in SVD was further demonstrated by the first three workload sizes of the SVD2 on a general $n \times n$ matrix (Figure~\ref{fig:svd2}). 
Dask (EC2) was faster than {\proj} for relatively smaller problem sizes, since the statically-deployed Dask {\small\texttt{distributed}} cluster supports direct worker-to-worker communication with less network I/O overhead (especially for large intermediate results), and the CPU resources of the cluster did not yet become a bottleneck.
Additionally, Dask (Laptop) suffered from OOM errors in the $50k \times 50k$ case and was unable to complete the workload. 
Finally, \proj executed the $100k \times 100k$ workload $3.1\times$ faster than Dask (EC2), 
again because of the elasticity of {\proj}. {\proj} does not require extra administration effort for scaling out the computation capacity, whereas Dask (EC2) would do, thus imposing extra burden to the end users. The number of Lambda functions used for each of the workloads was 84, 480, 295, and 1082, respectively. The $50k \times 50k$ workload used less Lambdas than the $25k \times 25k$ workload due to the strategy used to partition the initial input data. Different input data partitioning strategies may introduce different parallelism-communication tradeoffs and affect scalability. We plan to investigate partitioning strategies as part of our future work.

\begin{figure}[t]
\begin{center}
\includegraphics[width=0.4\textwidth]{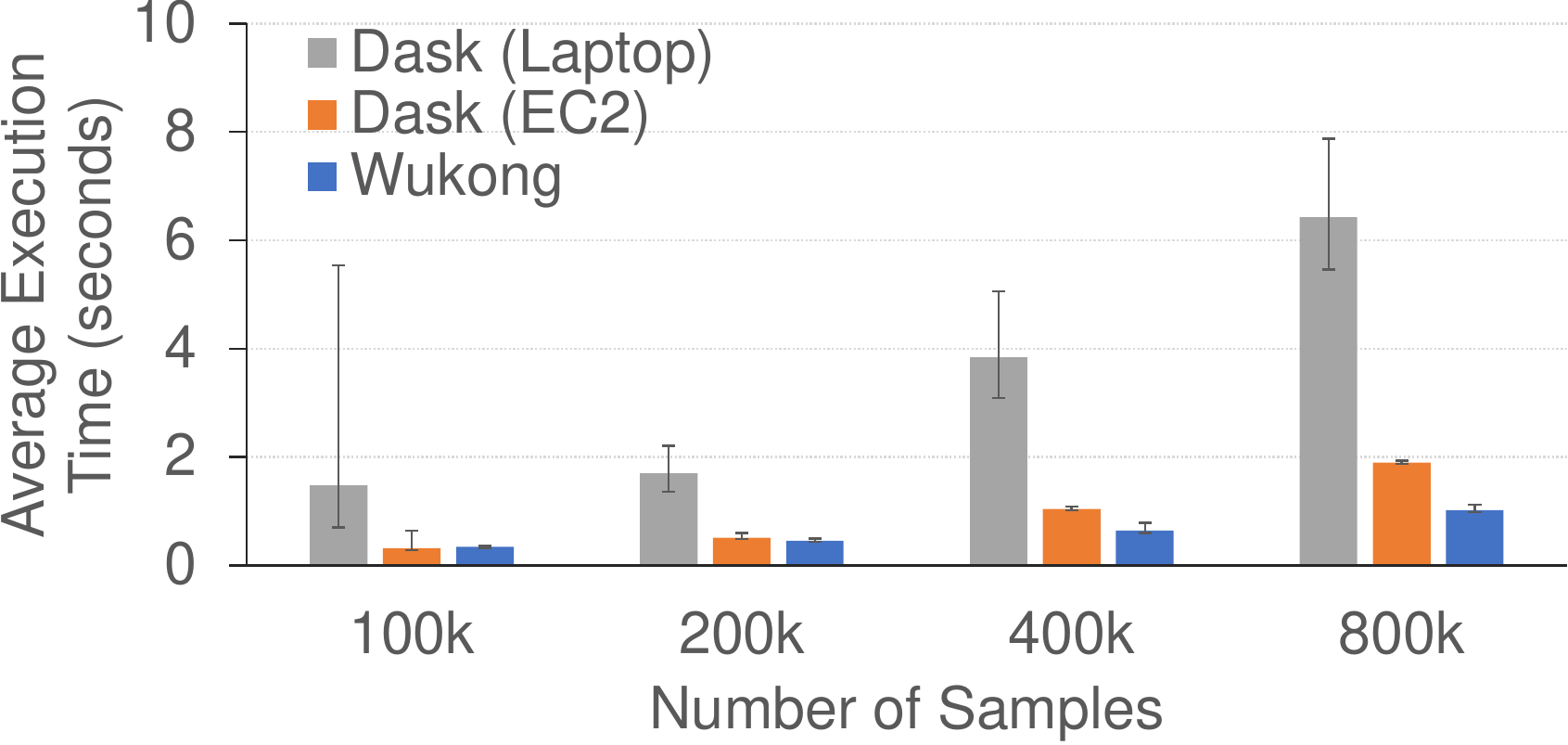}
\vspace{-5pt}
\caption{Performance comparison of SVC machine learning classification.}
\vspace{-15pt}
\label{fig:svc}
\end{center}
\end{figure}

Finally, we analyze the performance of SVC on \proj (Figure~\ref{fig:svc}). We varied the SVC problem size (in this case, number of samples) over the values $100k$, $200k$, $400k$, and $800k$. While Dask (EC2) completes the job slightly faster than \proj for the smallest problem size, the performance of \proj begins to exceed Dask (EC2) as the problem size increases. The performance gap increases as the problem size varied from $400k$ to $800k$. For a sample number of $800k$, \proj is able to execute the workload nearly $2\times$ as fast as Dask (EC2). This again strengthens our confidence that {\proj} can serve as a generic DAG engine for accelerating complex real-world applications such as machine learning.

\vspace{-5pt}
\subsection{Factor Analysis}
\label{subsec:factor}
\vspace{-3pt}

\begin{figure}[t]
\begin{center}
\includegraphics[width=0.46\textwidth]{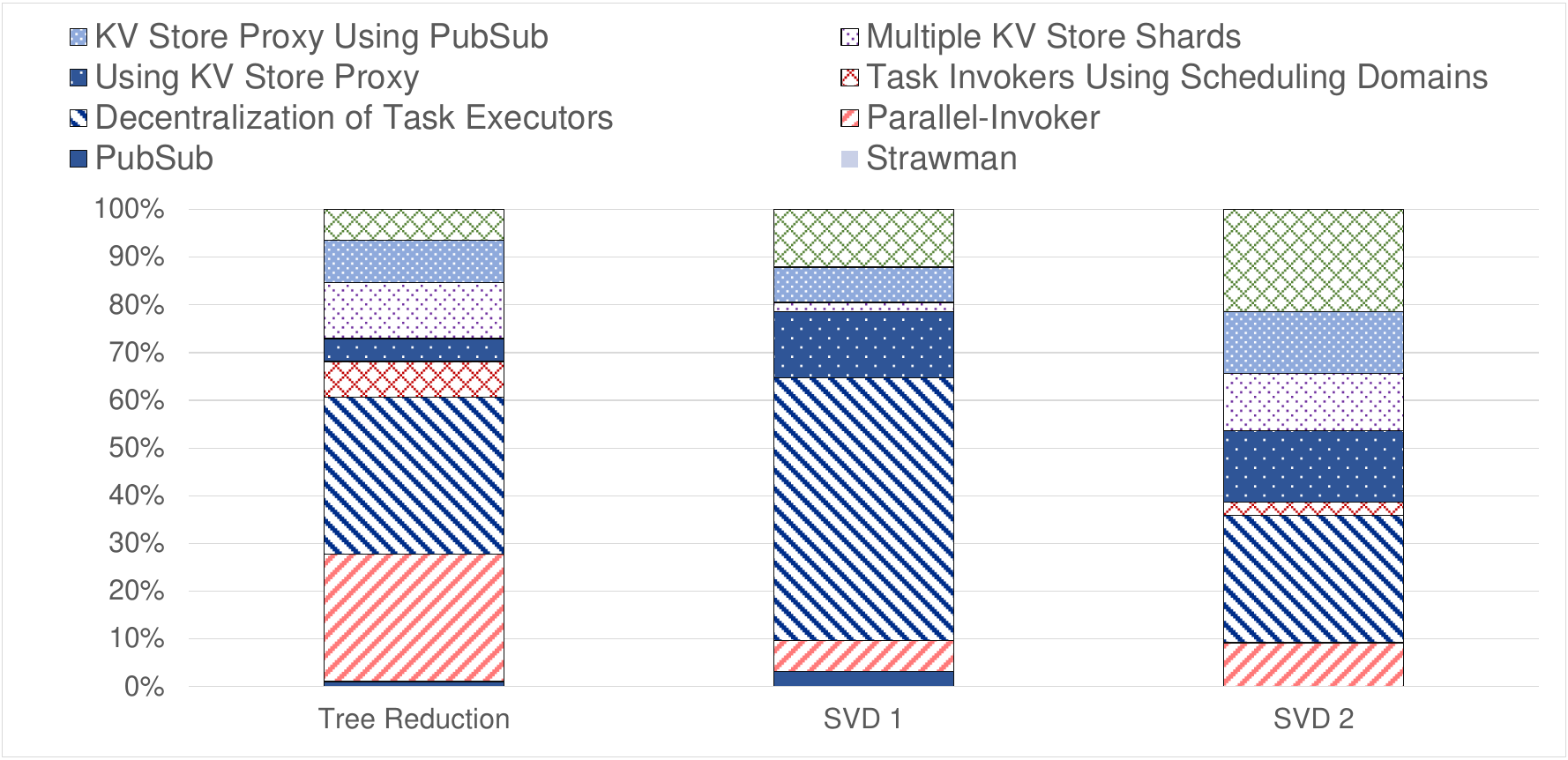}
\vspace{-5pt}
\caption{Contributing factors of different optimization techniques employed in {\proj}.}
\vspace{-20pt}
\label{fig:factor}
\end{center}
\end{figure}

\proj is able to effectively scale out to support large problem sizes and workloads. Figure~\ref{fig:factor} shows the amount that each major version of \proj contributed to the overall performance improvement from the original \textit{Strawman} version to the current version. The most significant improvement came as a result of the decentralization of the Task Executors. Prior to Task Executor decentralization, Task Executors would only execute the task initially given to them by the static scheduler. Once decentralized, Task Executors instead retrieved new tasks from the KV Store each time they completed the execution of their current task.  

Other significant improvements to the overall performance of \proj included the use of dedicated task invoker processes, which originated in the \textit{Parallel Invokers} version, and the use of the KV Store Proxy to parallelize large task fan-outs. The effect of the KV Store Proxy varied depending on the workload since workloads that lacked high fan-outs would not actually utilize the proxy. Switching the communication protocol used by the KV Store Proxy from TCP to Redis PubSub also resulted in a fairly substantial performance improvements. Just as for the static scheduler, Redis PubSub enabled the KV Store Proxy to handle a higher volume of messages from Task Executors. Finally, running each KV Store shard on its own separate VM resulted in a significant performance improvement. Initially, all KV Store shards were running on the same VM, which resulted in resource contention for network bandwidth. Placing each shard on its own VM eliminated this bottleneck.

\vspace{-5pt}
\subsection{Overhead Quantification}
\label{subsec:overhead}
\vspace{-3pt}

The overhead associated with storing and retrieving large intermediate data values during workload execution is a major factor that impacts {\proj}'s performance. For workloads characterized by short tasks and large communication overheads, \proj is not able to outperform Dask (EC2). This is most prevalent in the tree reduction workload without sleep delays, shown in Figure~\ref{fig:tr}, and when computing the SVD of a square matrix, shown in Figure~\ref{fig:svd2}. 

\begin{figure}[t]
\begin{center}
\includegraphics[width=0.5\textwidth]{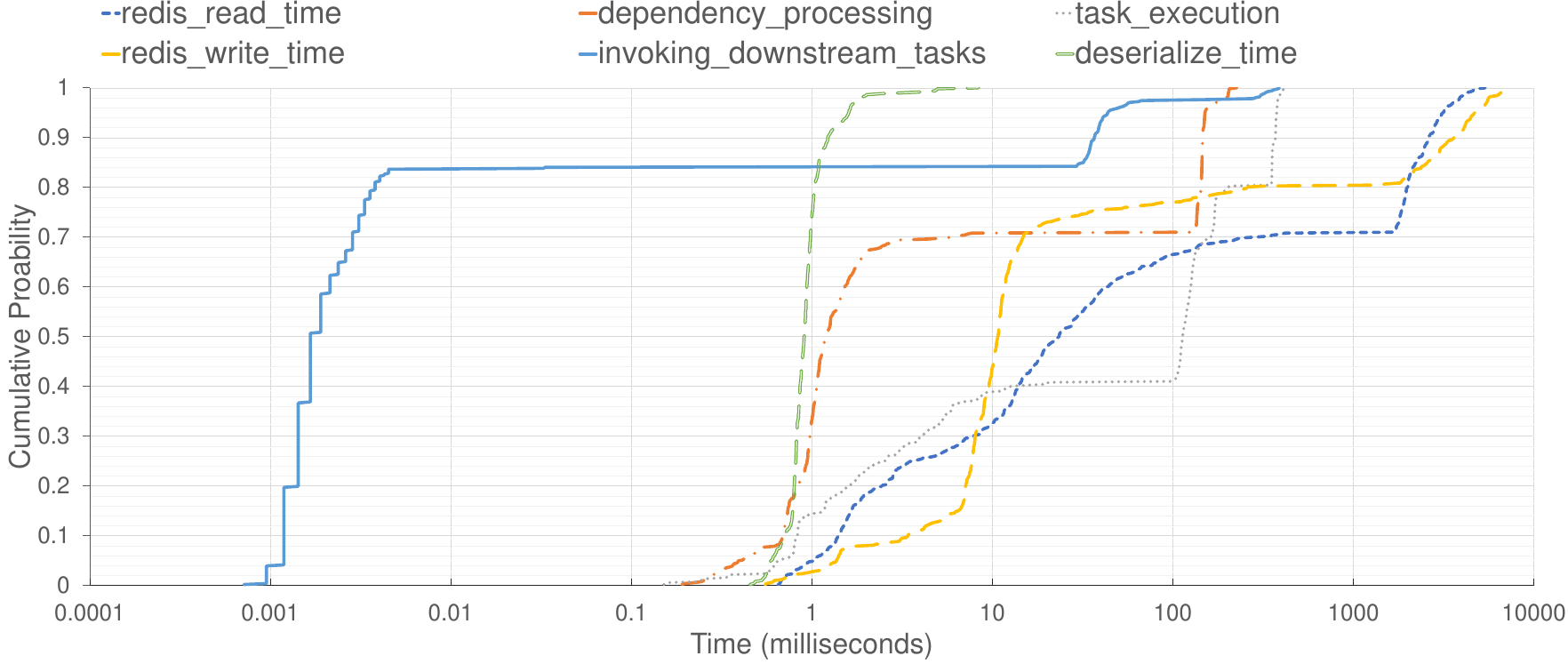}
\vspace{-15pt}
\caption{CDF breakdown of tasks in SVD2 with a $50k \times 50k$ matrix.}
\vspace{-20pt}
\label{fig:svd2_cdf}
\end{center}
\end{figure}

To quantify such I/O overhead, we conduct a detailed analysis with SVD2, by breaking down {\proj}'s execution duration into fine-grained factors. Figure~\ref{fig:svd2_cdf} shows a latency distribution of individual tasks in SVD2 of a $50k \times 50k$ matrix. We observe that there were a small number of KV store read and write operations which took upwards of ten seconds to complete. While a majority of tasks did not experience such communication overhead, the long network I/Os experienced by a minority of the tasks have a large impact on the workload's overall performance. 

In order to estimate the improvement in performance that \proj could obtain if we were to use an ideally-fast (i.e., fully-optimized) intermediate storage, we executed a modified variant of SVD2 in which all array data was randomly generated each time it was used (instead of being written in and retrieved from the KV store). In Figure~\ref{fig:svd2}, the right-most (yellow-colored) bar shows the performance of \proj with this ideal intermediate storage. While the performance of Dask (EC2) is still better than \proj (with idea storage) for the smallest workload size, the performance is roughly the same in the $25k \times 25k$ case. Moreover, \proj (with ideal storage) is able to perform $1.67\times$ faster than Dask (EC2) for the $50k \times 50k$ workload in this experiment. As discussed earlier, \proj (in its current form) is already able to outperform Dask (EC2) by over 115 seconds on-average for the largest problem size. When using an ideal KV store, \proj would execute the workload in $95.50\%$ less time than Dask (EC2). These results highlight the magnitude by which network communication overhead negatively affects the overall performance of \proj. 

Data locality is another key factor which influences the performance of \proj. Increased data locality enables Task Executors to carry out the workload without needing to retrieve dependent inputs from the KV Store. This reduces the communication overhead, thereby increasing the performance of the framework. The overall effect of data locality largely depends on the size of the data values kept in the Task Executor's local storage. Our analysis of the network I/O performance found that the transfer of intermediate data objects that were tens to hundreds of megabytes in size were the cause of longer execution times (as opposed to smaller intermediate data objects). Consequently, \proj is able to utilize the local data stores on Task Executors most effectively when large objects are stored. 

\vspace{-5pt}
\subsection{Limitations}
\label{subsec:overhead}
\vspace{-3pt}

One limitation of our evaluation is that we did not compare \proj against other serverless DAG engines. This was because all of the systems use different representations for their DAG's, and these representations are large and complicated. Consequently, it is nontrivial to convert a DAG from one system to another. A comparison between serverful Dask and \proj is possible because they use the same DAG representation. We are currently investigating DAG representations of other serverless DAG engines so that we can make a thorough comparison between {\proj} and other frameworks and understand the pros and cons of their design decisions.

\pdfoutput=1

\vspace{-6pt}
\section{Conclusion}
\label{sec:conclusion}
\vspace{-4pt}

We have presented {\proj}, a high-performance DAG engine that
implements decentralized scheduling by exploiting the elasticity and scalability of AWS Lambda to reduce network I/O overhead and improve data locality.
Our evaluation shows \proj is competitive with a traditional serverful DAG scheduler Dask
and demonstrates that decentralizing task scheduling 
contributes a significant portion 
of the improvement in overall performance.
As part of our future work, we are exploring new techniques to fundamentally improve the performance of intermediate storage for serverless DAG workloads. 

{\proj} is open sourced and is available at:
\url{https://github.com/mason-leap-lab/Wukong}.

\myparagraph{Acknowledgments.}
This work is sponsored in part by NSF under CCF-1919075 and an AWS Cloud Research Grant.

\balance
{
\footnotesize
\bibliographystyle{acm}
\bibliography{refs}
}
\newpage

\pdfoutput=1

\begin{appendices}

\section{Reproducibility Instructions}
\label{sec:Reproducibility}

The required components for running \proj are:
\begin{itemize}
\item Virtual Private Cloud (VPC)
\item AWS Lambda
\item EC2 Instances (Static Scheduler and KV Store)
\end{itemize}

This Appendix is organized as follows:
\begin{itemize}
\item Configuring the Virtual Private Cloud
\item Configuring the AWS Lambda Instance 
\item Configuring the EC2 Instances
\end{itemize}

\subsection{Configuring the Virtual Private Cloud (VPC)}
A VPC must be configured as EC2 instances are required to run within a VPC. 

\myparagraph{Step 1 --- Create the VPC}

\begin{figure}[h]
\begin{center}
\includegraphics[width=0.5\textwidth]{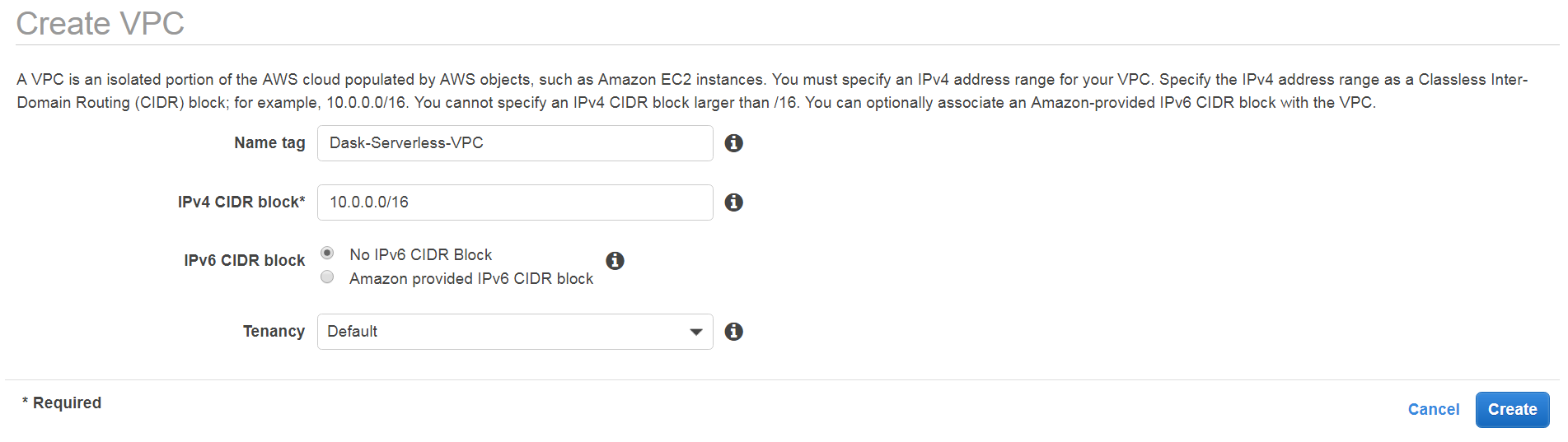}
\caption{Snapshot for creating VPC.}
\label{fig:vpc}
\end{center}
\end{figure}

Go to the VPC Console within AWS and click on ``Your VPCs'' (from the menu on the left-hand side of the screen). Next, click the blue ``Create VPC'' button. See Figure~\ref{fig:vpc} for a snapshot.

Provide a name tag and an IPv4 CIDR block. It is not necessary to provide an IPv6 CIDR Block. For Tenancy, select \textit{Default}. Once you have completed all of the fields, click the blue “Create” button. You should wait until the VPC has been created before continuing.

\myparagraph{Step 2 --- Create Two Security Groups}

\begin{figure}[h]
\begin{center}
\includegraphics[width=0.5\textwidth]{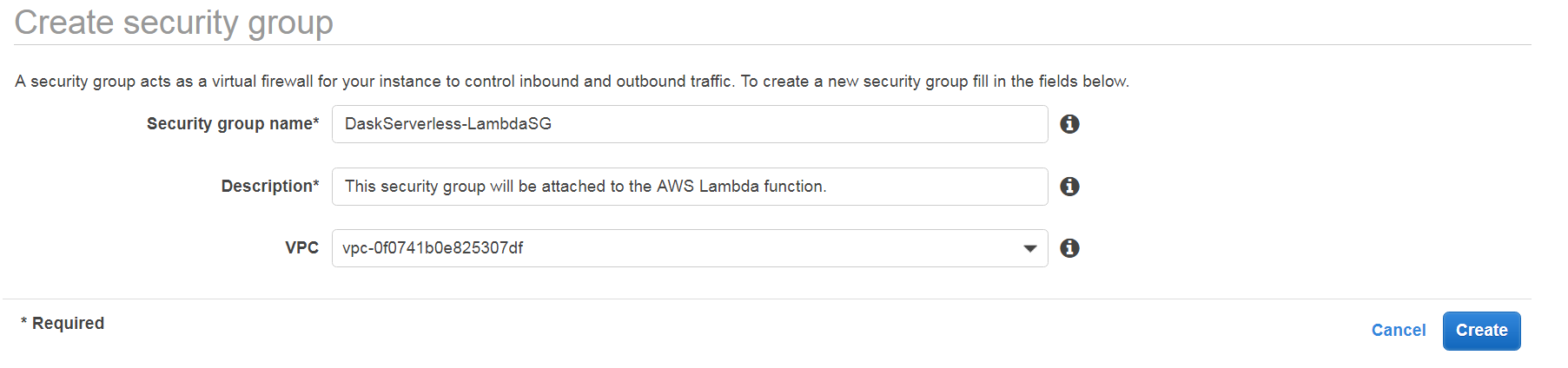}
\caption{Snapshot for creating security group.}
\label{fig:securityGroup}
\end{center}
\end{figure}

Next, create two security groups - one security group for the AWS Lambda function and another security group for the EC2 instances. See Figure~\ref{fig:securityGroup} for a snapshot.

From the VPC Console, select “Security Groups” from the menu on the left-hand side of the screen. Then click the blue “Create Security Group” button.

You will need to provide a security group name and a description. In the VPC field, select the VPC you created above. See Figure~\ref{fig:securityGroup}.

\myparagraph{Step 3 --- Configuring Inbound and Outbound Rules for the Security Groups}

Next, configure the inbound and outbound rules for the two security groups. This ensures that the different components will be able to communicate with one another (e.g. AWS Lambda and the KV Store). Note that you may use different ports then what we show here.

We first show a sample configuration for the AWS Lambda security group (i.e. the security group created for the AWS Lambda function). 

Figure~\ref{fig:inbound} shows a snapshot of the inbound rules:

\begin{figure}[h]
\begin{center}
\includegraphics[width=0.5\textwidth]{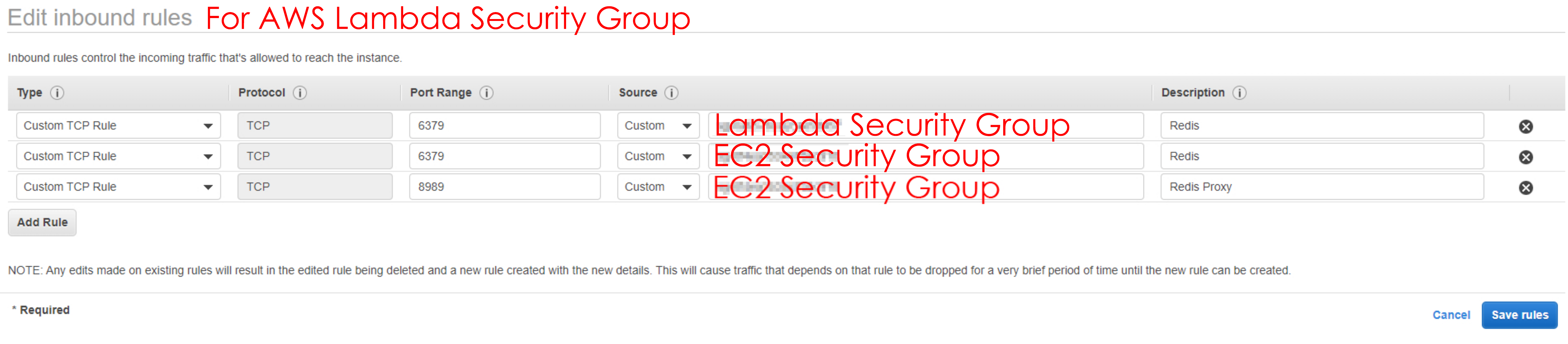}
\caption{Lambda's inbound rules.}
\label{fig:inbound}
\end{center}
\end{figure}

Figure~\ref{fig:outbound} shows a snapshot of the outbound rules:

\begin{figure}[h]
\begin{center}
\includegraphics[width=0.5\textwidth]{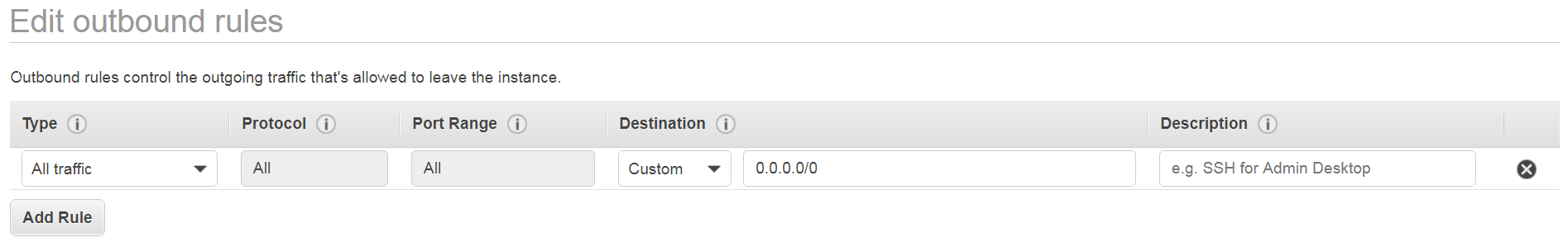}
\caption{Lambda's outbound rules.}
\label{fig:outbound}
\end{center}
\end{figure}

Here are sample configurations of the Inbound and Outbound rule configurations for the other security group.

Figure~\ref{fig:inboundSecurity} shows a snapshot of the inbound rules:

\begin{figure}[h]
\begin{center}
\includegraphics[width=0.5\textwidth]{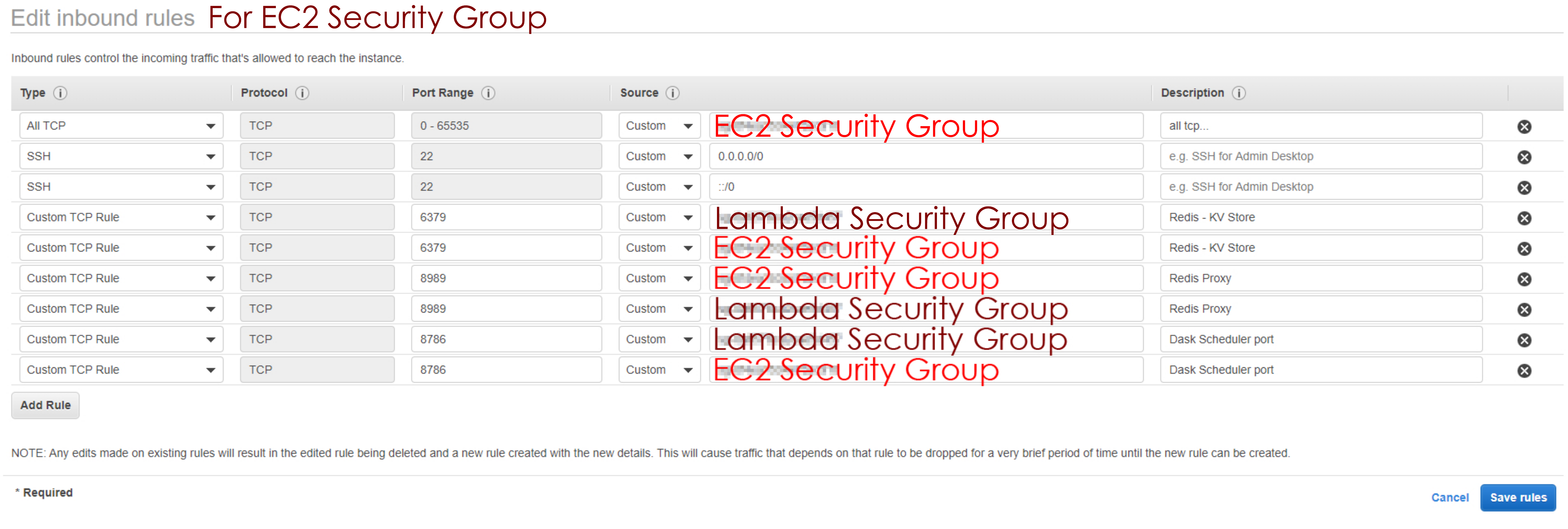}
\caption{How to edit the inbound rules under EC2 security group.}
\label{fig:inboundSecurity}
\end{center}
\end{figure}

Figure~\ref{fig:outbound2} shows a snapshot of the outbound rules:

\begin{figure}[h]
\begin{center}
\includegraphics[width=0.5\textwidth]{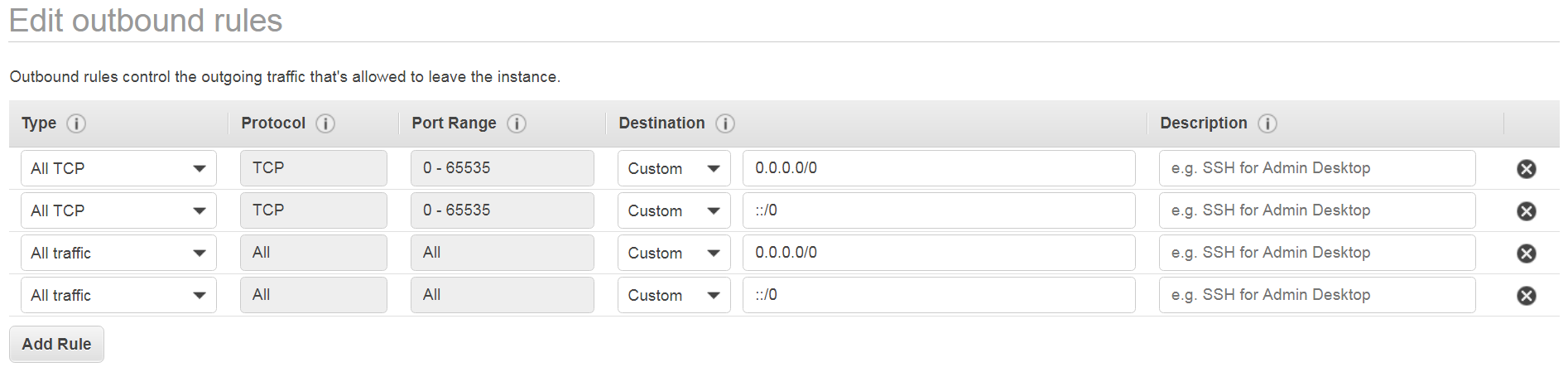}
\caption{Lambda's outbound rules (2).}
\label{fig:outbound2}
\end{center}
\end{figure}

\begin{figure}[h]
\begin{center}
\includegraphics[width=0.5\textwidth]{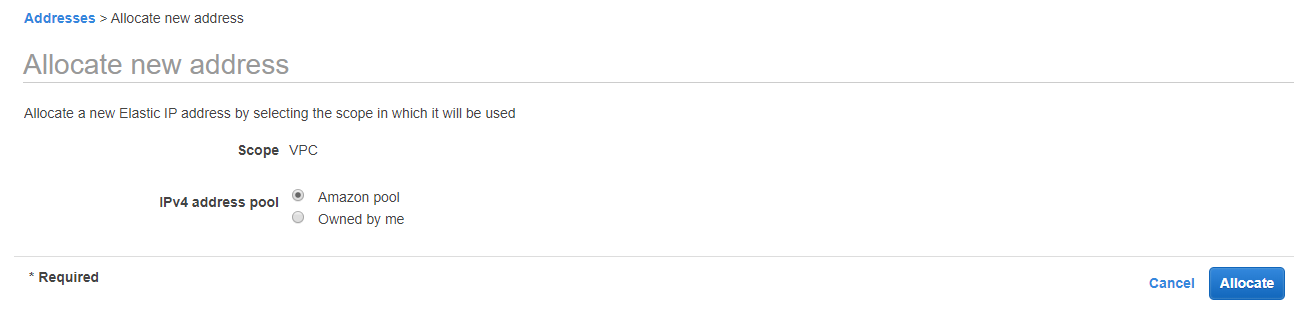}
\caption{Allocating a new elastic IP address.}
\label{fig:newAddress}
\end{center}
\end{figure}

\myparagraph{Step 4 - Allocate an Elastic IP Address }
Next, allocate an Elastic IP Address. From the VPC console, select ``Elastic IPs'' and click the blue ``Allocate new address'' button.
See Figure~\ref{fig:newAddress} as an example.

\begin{figure}[h]
\begin{center}
\includegraphics[width=0.5\textwidth]{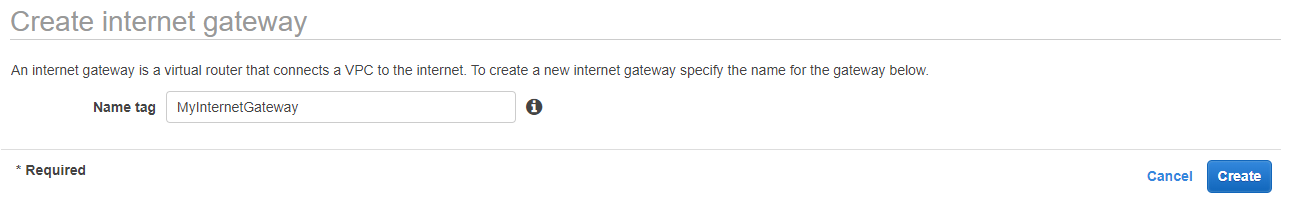}
\caption{Creating Internet gateway.}
\label{fig:gateway}
\end{center}
\end{figure}

\myparagraph{Step 5 --- Create Internet Gateway }
From the VPC console, select ``Internet Gateways'' and click the blue ``Create internet gateway'' button. Supply a name tag and click ``Create''. See Figure~\ref{fig:gateway} as an example.

\myparagraph{Step 6 --- Create Subnets for your VPC}
This important step is to create subnets for your VPC. For each availability zone,create a subnet specifically for the AWS Lambda function. When you create subnets, you will need to specify IPv4 CIDR blocks. 

As you will see in the screenshot of the subnet creation interface, two /16 IPv4 CIDR blocks were allocated to the VPC. This allowed the creation of a subnet containing a very large number of IP addresses for the AWS Lambda functions. 

\begin{figure}[h]
\begin{center}
\includegraphics[width=0.5\textwidth]{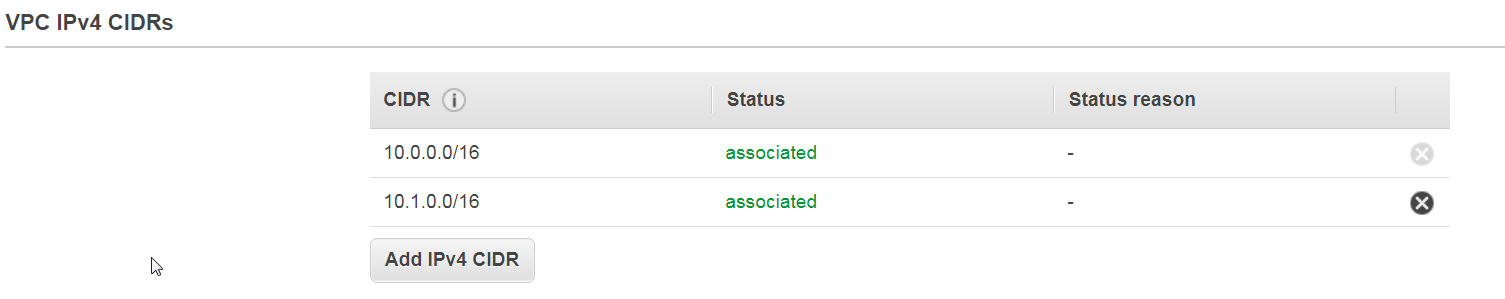}
\caption{Allocating more IPv4 CIDR blocks.}
\label{fig:ipv4}
\end{center}
\end{figure}

If you want to allocate more IPv4 CIDR blocks to your VPC, then simply go to the VPC console and select the VPC from the list of VPCs (under the ‘Your VPCs’ menu option). Click the white “Actions” button, then select ``Edit CIDRs''. At the bottom, you will see a table with a white ``Add IPv4 CIDR'' button underneath it. Click that button and specify the CIDR block. You can use the configuration shown in Figure~\ref{fig:ipv4} as a reference. 

To create a subnet, first go to the VPC console. Then select ``Subnets'' from the menu on the left. Finally, click the blue ``Create subnet'' button. 

Each subnet created for the Lambda functions, can be given a name tag like ``Lambda-Subnet-X'', where X is some number. The first subnet we created was ``Lambda-Subnet-1'', the second was ``Lambda-Subnet-2'', etc. You can choose your own names. For the VPC field, select the VPC you created earlier. 

We initially provisioned the CIDR blocks in chunks of 256 addresses. To do this, we specified the CIDR blocks as 10.0.2.0/24 for the first subnet. (The third number is 2 because 0 and 1 were taken by default subnets). We used 10.0.3.0/24 and 10.0.4.0/24 for the second and third subnets respectively. Again, this was done for all the availability zones. 

\begin{figure}[h]
\begin{center}
\includegraphics[width=0.5\textwidth]{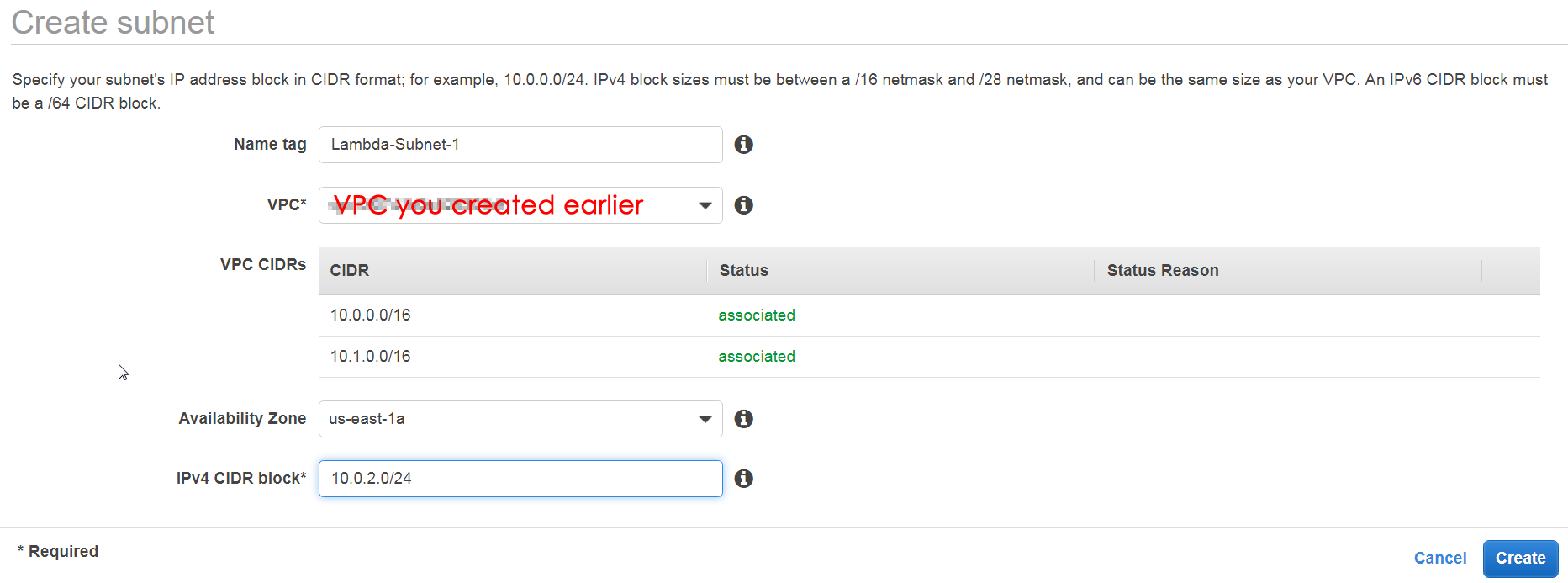}
\caption{Creating subnet.}
\label{fig:subnet}
\end{center}
\end{figure}

As I mentioned, we eventually allocated an additional CIDR block to the VPC (10.1.0.0/16), created a final Lambda subnet (in an arbitrary availability zone), and gave it the entire block, which is roughly 64,000 IP addresses. See Figure~\ref{fig:subnet} shown as below.

After you create however many subnets for your Lambda function, you then need to create a subnet for your KV Store and EC2 instances. Follow the same procedure as above, though I would give it a name that indicates the purpose of the subnet (i.e. that it is for the EC2/KV Store and not Lambda). 

\myparagraph{Step 7 --- Edit the Route Tables \& Creating a NAT Gateway}
We created two route tables. One of the route tables was for the subnets created for EC2 and the KV Store. The other route table was for subnets created for AWS Lambda.

\begin{figure}[h]
\begin{center}
\includegraphics[width=0.5\textwidth]{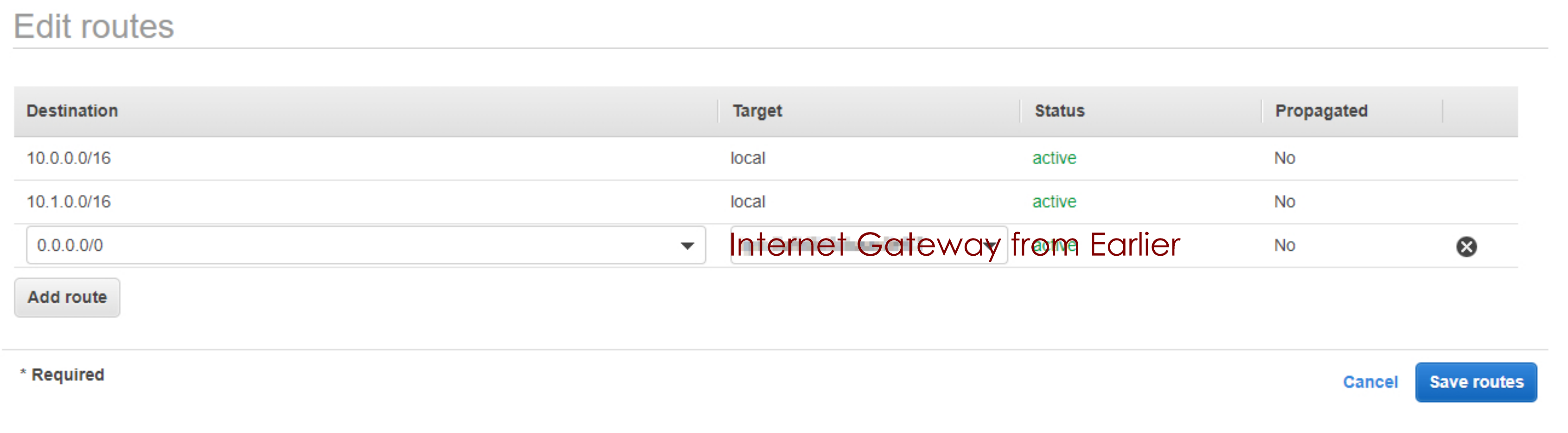}
\caption{Editing routes.}
\label{fig:routes}
\end{center}
\end{figure}

From the VPC Console, select ``Route Tables'' and then click the blue ``Created route table'' button (Figure~\ref{fig:routes}). We first create the route table to be used with public subnets (EC2 and KV Store subnets). 

\myparagraph{Step 8 --- KV Store/EC2 Route Table}

\myparagraph{Step 9 --- Create a NAT Gateway}
Next, create a NAT Gateway. A NAT Gateway is placed into a public VPC subnet to enable outbound internet traffic from instances in a private subnet. WE placed this NAT Gateway in one of the subnets created for the EC2/KV Store. 

\myparagraph{Step 10 --- AWS Lambda Route Table}
A route table must be configured to give AWS Lambda access to the Internet (since it is restricted when Lambda functions are running within a VPC).

\begin{figure}[h]
\begin{center}
\includegraphics[width=0.5\textwidth]{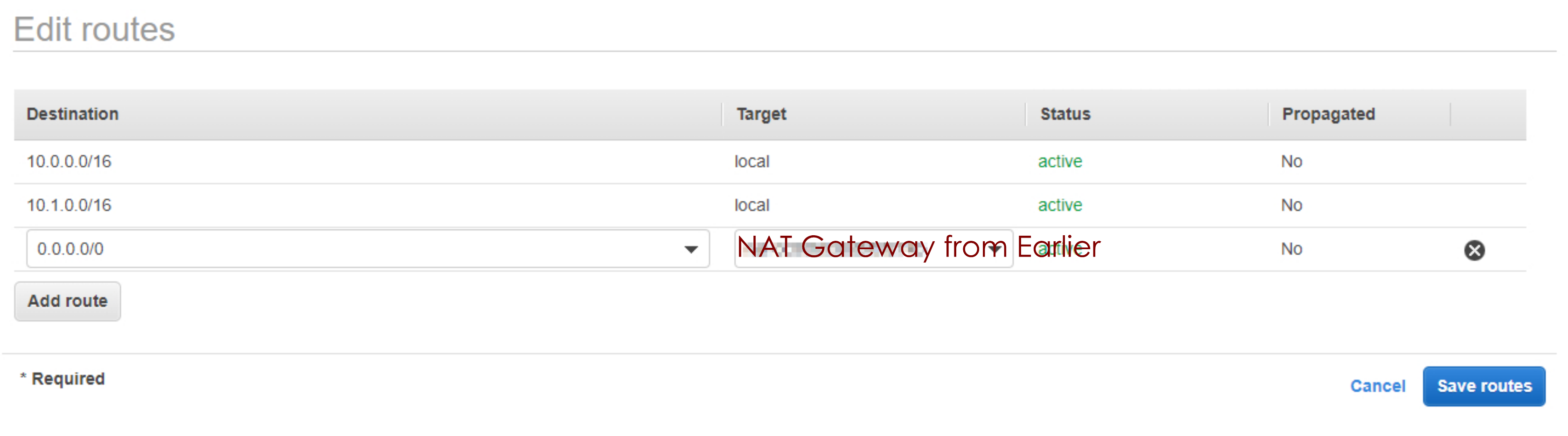}
\caption{Editing routes with NAT gateway.}
\label{fig:routesNAT}
\end{center}
\end{figure}

Click on the ``Route Tables'' option from the menu on the left of the VPC Dashboard. Click the blue ``Create route table'' button. Provide a name tag and select the VPC you created earlier. Next, select this newly created route table from the list of route tables (from the “Route Tables” section of the VPC Dashboard) and select the ``Routes'' tab. The various tabs will be at the bottom of your screen.
Click the white ``Edit routes'' button and fill in the routes like so. Select the NAT Gateway you just created (Figure~\ref{fig:routesNAT}).

Modify the Subnet Associations for this route table. Select the newly created route table from the list of route tables (which is found under the ‘Route Tables’ menu option in the VPC console). Look for the “Subnet Associations” tab at the bottom of your screen and click the white “Edit subnet associations” button. Select all subnets that were created for your AWS Lambda function.

\begin{figure}[h]
\begin{center}
\includegraphics[width=0.5\textwidth]{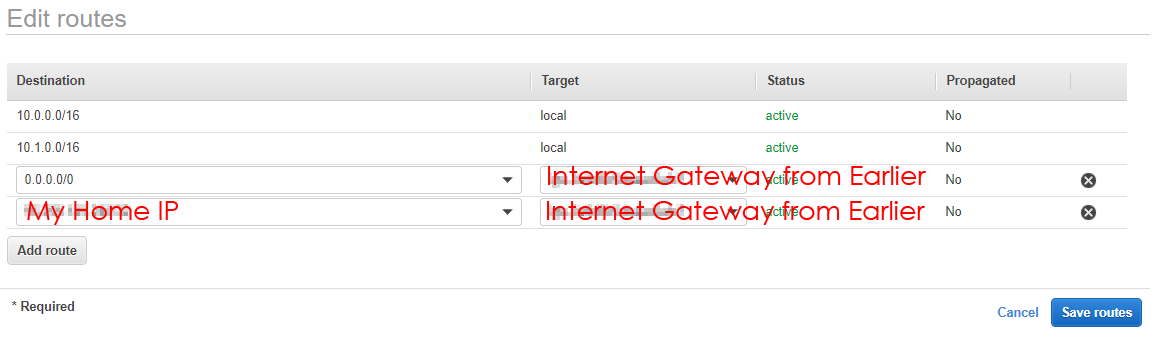}
\caption{Editing routes with Internet gateway.}
\label{fig:routesIG}
\end{center}
\end{figure}

\myparagraph{Step 11 --- EC2/KV Store Routing Table}
Create a second route table. Fill in the routes as follows. Use the Internet Gateway created earlier. You do not need the ``Home IP'' route. See Figure~\ref{fig:routesIG} as a reference.

\myparagraph{Step 12 - Downloading the Source Code}
The source code for the Static Scheduler, KV Store Proxy, and AWS
Lambda Task Executor can be cloned or downloaded from
\url{https://github.com/mason-leap-lab/Wukong}.

\subsection{Configuring the AWS Lambda Function (Task Executor)}

Create a new AWS Lambda function. You can name it whatever you want, but you will need to change the name of the function in the source code to match whatever you specify.

Once you have downloaded the source code for the Lambda function, package the contents in a .ZIP file and upload the file to the AWS Lambda function. 

There are a number of dependencies required by the AWS Lambda function. These can be bundled with the source code, or you can place them in a Lambda Layer and add the layer to your function. Two of the dependencies include \textit{SciPy} and \textit{NumPy}. Amazon provides a unique Lambda Layer for these libraries, and it is recommended to use that layer when adding the dependencies to your Lambda function. The full list of dependencies required for the Lambda function is as follows:
\begin{enumerate}
\item SciPy
\item NumPy
\item Dask
\item aws\_xray\_sdk 
\item cloudpickle
\item dateutil
\item docutils
\item future
\item jmespath
\item jsonpickle
\item libfuturize
\item libpasteurize
\item lz4
\item msgpack
\item pandas
\item past
\item pytz
\item redis
\item tlz
\item toolz
\item tornado
\item uhashring
\item urllib3
\item wrapt
\end{enumerate}

If you would like to be able to execute Dask-ML (machine learning) workloads, then you also need the following dependencies: 
\begin{enumerate}
\item attr
\item click
\item dask\_glm
\item dask\_ml
\item joblib
\item multipledispatch
\item numba
\item packaging
\item psutil
\item sklearn
\item sortedcontainers
\item tblib
\item yaml
\item zict
\end{enumerate}

There are a number of changes you must make to the AWS Lambda function's default configuration. First of all, the \textit{Handler} field should be set to ``function.lambda\_handler''. It is recommended to allocate at least 1986MB of memory to the Function, if not the full 3GB. For \textit{Timeout}, we recommend setting the timeout to at least two minutes. 

\begin{figure}[h]
\begin{center}
\includegraphics[width=0.5\textwidth]{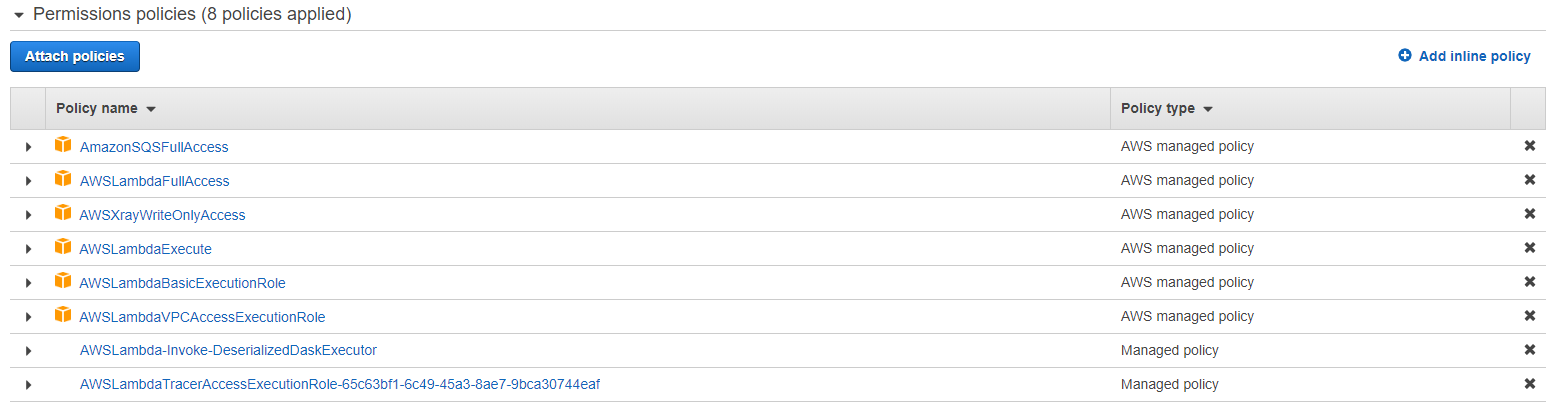}
\caption{IAM policies.}
\label{fig:iam}
\end{center}
\end{figure}

The \textit{Execution role} for the Lambda function needs certain policies attached to it. The required policies are shown in Figure~\ref{fig:iam}. You should create an IAM role with the policies shown, and assign that role to the AWS Lambda function. 

Under the \textit{Network} section of the configuration, specify the VPC you created earlier for the ``Virtual Private Cloud'' field. Select all of the subnets you created for the Lambda function, and select the security group you created as well. The Lambda function has X-Ray tracing built into it so it is recommended to enable "Active tracing" in the \textit{AWS X-Ray} section of the function configuration. 

\subsection{Executing \proj }
In order to execute \proj, you will need one EC2 VM for the Static Scheduler and at least one additional VM to host the KV Store Proxy and a KV Store server. 

Using the method of your choice, launch the desired number of EC2 virtual machines ($\geq$ 2). When configuring the machines, make sure to place them in the VPC you created earlier. In the EC2 Console, this is done using the drop-down menu labeled \textbf{Network} in \textit{Step 3. Configure Instance Details} of the Launch Instance Wizard. Select the subnets you created earlier for the Static Scheduler and KV Store shards. 

Also, assign to the instances the security group created earlier. You should set the \textit{Auto Assign IPv4 Address} property to \textit{True}.

When it comes time to actually launch all of the components, the Redis shards should be launched first. Once they are running, the KV Store Proxy should be launched next. Finally, the Static Scheduler should be launched. 

Additionally, if you would like to use a name other than the default name given in this Appendix for the Lambda function, you must update this value in the AWS Lambda function's code, the KV Store Proxy, and the Static Scheduler. In the case of the Lambda function, modify the value for the \textit{lambda\_function\_name} variable found on line 56 of \textbf{function.py} at the time of writing this. This change should be made in the KV Store Proxy as well as the Static Scheduler. 

\myparagraph{Redis}
Once your instances have launched and are fully initialized, you must start the various components of \proj. First, SSH into the instances you intend to use for KV Store shards. If you are not using an AMI that has Redis pre-installed, then install Redis using your method of choice. You should also create a Redis server configuration file with the following contents:\\\\
\noindent\fbox{%
    \parbox{\linewidth}{%
        protected-mode no\\
        bind 0.0.0.0\\
        port 6379\\
        appendonly no\\
        save ''"
    }%
}\\

You may change the port to whatever you'd like; but keep track of what port you use for each server if they all use different ports. 

Once your config file is created, you may start the Redis server as follows:\\\\
\noindent\fbox{%
    \parbox{\linewidth}{%
        redis-server /path/to/configuration/file/redis\_config.conf
    }%
}\\

\myparagraph{KV Store Proxy}
Generally speaking, you will want to execute the KV Store Proxy process co-located with one of the KV Store shards (although this is not strictly required; it will function with or without co-location). 

The proxy was written and tested with Python3. The required dependencies are:

\begin{enumerate}
\item toolz
\item tornado
\item cloudpickle
\item six
\item aioredis
\item asyncio
\item redis-py
\item boto3
\item uhashring
\item dask
\item msgpack
\end{enumerate}

These dependencies can generally be installed using pip. See the dependencies' respective websites for specific installation instructions. 

If you need to change the name of the AWS Lambda function invoked by the Proxy, you should pass a value for the keyword argument ``function\_name'' to the \textit{ProxyLambdaInvoker} constructor found on line 377 of \textbf{proxy.py} at the time of writing this. This change should also be made in the AWS Lambda function itself as well as in the static scheduler. 

One the source code has been downloaded and the required dependencies have been installed, go to the \textit{RedisProxy} directory found in the KV Store Proxy source code. The syntax for executing the Proxy is as follows:\\\\
\noindent\fbox{%
    \parbox{\linewidth}{%
        python3 proxy.py -res ``redis-endpoint-1'' ``redis-endpoint-2'' ... ``redis-endpoint-\textit{n}'' -rps 1 2 3 ... \textit{n}
    }%
}\\

If all KV Store Shards are listening on the same port, then you only need to specify that port once for the \textit{-rps} argument. Otherwise, the first port specified corresponds to the first endpoint specified; the second port specified corresponds to the second endpoint specified; and so on. Note that you must use ports that you have configured in the Inbound and Outbound rules of the security groups created earlier.

\textbf{Example \#1:}\\
Redis Shard 1: ``ec2-abc-123-xyz:6379''\\
Redis Shard 2: ``ec2-abc-321-xyz:6380''\\

\noindent\fbox{%
    \parbox{\linewidth}{%
        python3 proxy.py -res ``ec2-abc-123-xyz'' ``ec2-abc-321-xyz'' -rps 6379 6380
    }%
}\\

\textbf{Example \#2:}\\
Redis Shard 1: ''ec2-abc-123-xyz:6379"\\
Redis Shard 2: ''ec2-abc-321-xyz:6379"\\
Redis Shard 3: ''ec2-abc-312-xyz:6379"\\

\noindent\fbox{%
    \parbox{\linewidth}{%
        python3 proxy.py -res ``ec2-abc-123-xyz'' ``ec2-abc-321-xyz'' ``ec2-abc-312-xyz'' -rps 6379
    }%
}\\

\myparagraph{Static Scheduler}
The dependencies required for execution of the Static Scheduler include:
\begin{enumerate}
\item click $\geq$ 6.6
\item cloudpickle $\geq$ 0.2.2
\item dask $\geq$ 0.18.0
\item msgpack
\item psutil $\geq$ 5.0
\item six
\item sortedcontainers !=2.0.0, !=2.0.1
\item tblib
\item pytest 
\item moto 
\item mock
\item toolz $\geq$ 0.7.4
\item tornado $\geq$ 5
\item zict $\geq$ 0.1.3\\
\textit{Compatibility packages}
\item futures; python\_version $<$ '3.0'
\item singledispatch; python\_version $<$ '3.4'
\item pyyaml
\end{enumerate}

There are additional dependencies required for development. These dependencies include:
\begin{enumerate}
\item joblib $\geq$ 0.10.2
\item mock $\geq$ 2.0.0
\item pandas $\geq$ 0.19.2
\item numpy $\geq$ 1.11.0
\item bokeh $\geq$ 0.12.3
\item requests $\geq$ 2.12.4
\item pyzmq $\geq$ 16.0.2
\item ipython $\geq$ 5.0.0
\item jupyter\_client $\geq$ 4.4.0
\item ipykernel $\geq$ 4.5.2
\item pytest $\geq$ 3.0.5
\item prometheus\_client $\geq$ 0.6.0
\item jupyter-server-proxy $\geq$ 1.1.0
\end{enumerate}

If you need to change the name of the AWS Lambda function invoked by the Static Scheduler, pass a value for the keyword argument ``function\_name'' to the \textit{BatchedLambdaInvoker} constructor found on line 1356 of \textbf{scheduler.py} at the time of writing this. This change should also be made in the AWS Lambda function itself as well as the KV Store Proxy. 

Once all dependencies are installed and the source code is downloaded, navigate to the \textit{distributed} folder in the source code folder hierarchy. Start an interactive session with the Python interpreter by executing the "python" or "python3" command in your terminal session. 

There are a few key objects you need to create in order to execute workloads on \proj. These objects include an instance of \textit{LocalCluster} and \textit{Client}.

\definecolor{dkgreen}{rgb}{0,0.6,0}
\definecolor{gray}{rgb}{0.5,0.5,0.5}
\definecolor{mauve}{rgb}{0.58,0,0.82}

\lstset{frame=tb,
  language=Python,
  aboveskip=3mm,
  belowskip=3mm,
  showstringspaces=false,
  columns=flexible,
  basicstyle={\small\ttfamily},
  numbers=none,
  numberstyle=\tiny\color{gray},
  keywordstyle=\color{blue},
  commentstyle=\color{dkgreen},
  stringstyle=\color{mauve},
  breaklines=true,
  breakatwhitespace=true,
  tabsize=3
}

\begin{lstlisting}
from distributed import LocalCluster, Client
# The LocalCluster object will 
# create the Static Scheduler object.
local_cluster = LocalCluster(
                    host = "ip:port",
                    n_workers = 0,
                    proxy_address = "ip",
                    proxy_port = 8989,
                    redis_endpoints = [("ip":port), ("ip": port)],
                    num_lambda_invokers = 20, 
                    max_task_fanout = 10)

# Create an instance of Client. 
# This is how you interface with the 
# Static Scheduler.
client = Client(local_cluster)
\end{lstlisting}

The following is a discussion of the arguments to the \textit{LocalCluster} constructor. 
\begin{itemize}
\item \textit{host} should be specified in the format \textit{IP-Address}:\textit{Port}, where \textit{IP-Address} is the public IPv4 address of the EC2 instance on which you are going to execute the static scheduler. \textit{Port} should be 8786.
\item \textit{n\_workers} should simply be set to zero.
\item \textit{proxy\_address} should be the public IPv4 address of the EC2 instance on which you are executing the KV Store Proxy.
\item \textit{proxy\_port} should be the default port of 8989, though you can use a different port if you configured your security groups correctly earlier in the process.
\item \textit{redis\_endpoints} is a list of tuples of the form $[(ip_{1}, port_{1}), (ip_{2}, port_{2}), ..., (ip_{n}, port_{n})]$ where \textit{n} is the number of KV Store shards you have running. Each IP address is the public IPv4 address of the EC2 VM on which the KV Store shard is executing.  
\item \textit{num\_lambda\_invokers} is the number of leaf Task Invoker processes you want the Static Scheduler to create. 
\item \textit{max\_task\_fanout} is the threshold such that when a given task has \textit{max\_task\_fanout} or more downstream tasks, it will offload invocations to the KV Store Proxy.
\end{itemize}

Once you have executed the code shown above, you can interact with the Static Scheduler just as you would with traditional Dask/Dask Distributed. For example, the SVD computation for a tall-and-skinny matrix can be done as follows:

\begin{lstlisting}
import dask.array as da

X = da.random.random((200000, 100), 
            chunks = (10000, 100))
u, s, v = da.linalg.svd(X)
# This starts the computation.
v.compute()
\end{lstlisting}

\end{appendices}

\end{document}